\documentclass[aps,prd,superscriptaddress,amsmath,groupedaddress,twocolumn,final]{revtex4}

\usepackage{color,hangcaption,hhline,psfrag,rotating,amssymb}
\usepackage[hang,nooneline]{subfigure}
\usepackage[color]{showkeys}

\newcommand{\bea}{\begin{eqnarray}}
\newcommand{\eea}{\end{eqnarray}}
\newcommand{\beq}{\begin{equation}}
\newcommand{\eeq}{\end{equation}}
\newcommand{\del}{{\bf \Delta}}

\newcommand{\delv}{{\bf \del}}
\newcommand{\delfour}{{\Delta^{(4)}}}
\newcommand{\delsq}{\Delta^{(2)}}

\newcommand{\be}{\begin{equation}}
\newcommand{\ee}{\end{equation}}
\newcommand{\msb}{\overline{\mathrm{MS}}}

\newcommand{\mbare}{\mbox{$M_b^0$}}

\newcommand{\nl}{\nonumber \\}

\newcommand{\Ev}{{\bf E}}
\newcommand{\Bv}{{\bf B}}

\newcommand{\sigmav}{\mbox{\boldmath$\sigma$}}

\definecolor{gray}{rgb}{0.6,0.6,0.6}

\begin{document}
\title{The $\Upsilon$ spectrum and $m_b$ from full lattice QCD}

\author{A.Gray}
\email{agray@mps.ohio-state.edu}
\affiliation{Department of Physics, The Ohio State University, Columbus, OH 43210, USA}
\author{I. Allison}
\affiliation{Department of Physics and Astronomy, University of Glasgow, Glasgow, UK}
\author{C. T. H. Davies}
\email{c.davies@physics.gla.ac.uk}
\affiliation{Department of Physics and Astronomy, University of Glasgow, Glasgow, UK}
\author{E. Gulez}
\affiliation{Department of Physics, The Ohio State University, Columbus, OH 43210, USA}
\author{G. P. Lepage}
\affiliation{Laboratory of Elementary-Particle Physics, Cornell University, Ithaca, NY 14853}
\author{J. Shigemitsu}
\affiliation{Department of Physics, The Ohio State University, Columbus, OH 43210, USA}
\author{M. Wingate}
\affiliation{Institute for Nuclear Theory, University of Washington, Seattle, WA 98195, USA}

\collaboration{HPQCD and UKQCD collaborations}
\noaffiliation

\date{\today}

\begin{abstract}
We show results for the $\Upsilon$ spectrum calculated in lattice QCD 
including for the first time vacuum polarization effects 
for light $u$ and 
$d$ quarks as well as $s$ quarks. We use 
gluon field configurations generated by the MILC collaboration. The calculations 
compare the results for a variety of $u$ and $d$ quark masses, as well 
as making a comparison to quenched results (in which quark 
vacuum polarisation is ignored) and 
results with only $u$ and $d$ quarks. The $b$ quarks in the $\Upsilon$
are treated in lattice Nonrelativistic QCD through NLO in an expansion 
in the velocity of the $b$ quark. 
We concentrate on accurate results for orbital and radial splittings where we 
see clear agreement with experiment once $u$, $d$ and $s$ 
quark vacuum polarisation effects are included. 
This now allows a consistent determination 
of the parameters of QCD. We demonstrate this consistency 
through the agreement of the $\Upsilon$ and $B$ spectrum using 
the same lattice bare $b$ quark mass.
A one-loop matching to continuum QCD gives a value
for the $b$ quark mass in full lattice QCD for the first time. We obtain 
$m_b^{\overline{MS}}(m_b^{\overline{MS}})$ = 4.4(3) GeV. 
We are able to give physical results for the heavy quark 
potential parameters, $r_0$ = 0.469(7) fm and $r_1$ = 0.321(5) fm.
Results for the fine structure in the 
spectrum and the $\Upsilon$ leptonic width are also presented.
We predict the $\Upsilon - \eta_b$ splitting to be 61(14) MeV, 
the $\Upsilon^{\prime} - \eta_b^{\prime}$ splitting as 30(19) MeV and 
the splitting between the $h_b$ and the spin-average of the $\chi_b$ states 
to be less than 6 MeV. 
Improvements to these calculations that will be made in the near future are discussed.

\end{abstract}


\maketitle

\section{Introduction}\label{sec:intro}
The calculation of the spectrum of bottomonium and charmonium states is a key 
one for lattice QCD, for several reasons:
\begin{itemize}
\item 
There are many `gold-plated' (narrow, stable and experimentally 
well-characterised) states 
whose masses can be calculated accurately in lattice QCD. 
\item 
Splittings in the spectrum have particularly good properties to use 
for determining the scale of the theory, $\Lambda_{QCD}$. In 
lattice QCD determining this scale is equivalent to knowledge of 
the lattice spacing and it is an important first calculation that must 
be done before quoting other dimensionful 
quantities on the lattice such as hadron masses, or giving a scale 
to $\alpha_s$. 
\item
The bottomonium and charmonium calculations provide a test of the 
lattice QCD actions used for the heavy $b$ and $c$ quarks. These can then be 
used for the $B$ and $D$ physics results needed for the experimental 
programme charged with testing the consistency of CP violation in 
the Standard Model. Lattice QCD numbers are critical, for example,  to push 
down the limits on determination of the elements of the CKM matrix
to the few percent level.   
\item 
The control of systematic errors from the lattice for heavyonium masses is 
well developed and the results are statistically very 
precise. 
Vacuum polarisation effects for heavy quarks are not included 
because they are known perturbatively
to have very little effect. This makes the tuning 
of the heavy quark mass much simpler than that of 
light quark masses.  
The heavyonium spectrum is sensitive to 
$u$, $d$, and $s$ quark vacuum polarisation effects, but is not 
very sensitive to the actual values of the light quark masses, 
once they are low enough. 
All of these points make calculations of heavyonium masses 
in the presence of light 
quark vacuum polarization relatively simple, and a good test of lattice methods.
\end{itemize}

We need to achieve 
statistical and systematic errors of a few percent in order for lattice QCD 
to provide useful information for the particle physics programme 
not obtainable by other methods. 
New results~\cite{us,ourms,milc3,bc,dsemi} indicate that this is starting to be possible, 
building on many years of work by the lattice community to understand 
the sources of error in lattice calculations. The developments that 
have enabled this progress are: improved actions, the inclusion of 
light sea quarks (i.e. light quark vacuum polarization effects) and 
effective theories for handling heavy quarks. 
In particular the inclusion of $u$, $d$, and $s$ sea quarks 
with light $u$ and $d$ quark masses has 
produced agreement with experiment for a range of hadron 
masses for the first time~\cite{us}. A key component of this, for 
the reasons above, has been a set of results on the $\Upsilon$ spectrum
and we describe these in more detail here. 

Section~\ref{sec:calc}  describes details of the lattice calculations and 
section~\ref{sec:results}  gives the results obtained. In sections~\ref{sec:scale}  and~\ref{sec:radorb} 
we describe the analysis that allows the extraction of 
the lattice spacing and we compare the spectrum 
obtained to experiment. In section~\ref{sec:mb}  we describe the determination 
of the $b$ quark mass from the results. In sections~\ref{sec:fine} 
and~\ref{sec:lepwid} we discuss the fine structure in the spectrum and the 
determination of the leptonic width of the $\Upsilon$ and 
its radial excitations. 
We conclude in section~\ref{sec:conc}  with a discussion of further improvements that 
will be made, particularly to these last two calculations. 

\section{The lattice calculation}\label{sec:calc} 

The bottomonium spectrum is obtained by calculating correlators 
for bottomonium hadrons on configurations of gluon fields 
generated by a Monte Carlo procedure. The gluon field configurations 
in an ensemble are generated with a probability distribution given by 
$\exp(-S_{QCD})$ where $S_{QCD}$ is the QCD action. Averaging
over results on an ensemble then gives the Feynman Path Integral 
result for the bottomonium correlator. First we describe the 
different ensembles of gluon field configurations that 
we have used and then the method for calculating the bottomonium 
correlators. We will then estimate the size of systematic 
errors in the calculation and finally discuss the fitting procedure which 
enables us to obtain energies and amplitudes for different 
states from the correlators. 

The ensembles of gluon field configurations were generated by 
the MILC collaboration~\cite{milc1,milc2}. Different ensembles 
include the effect of zero, two and three flavors of sea quarks. 
Ensembles with zero flavors of sea quarks are called `quenched', 
and those with sea quarks included are called `unquenched' (or `dynamical'). 
The ensembles with three flavors are most interesting from a 
physical point of view because sea $u$, $d$, and $s$ 
quarks are an important component of the vacuum of QCD in 
the real world. The MILC collaboration has made ensembles 
with a range of masses for the sea $u$ and $d$ quark 
masses (which are taken to be equal) down to much lighter
values than before and within reach 
of their real world values. The ensembles in which the sea $u$ and 
$d$ quarks are much lighter than the sea $s$ quark will be 
called `2+1 flavor' ensembles. The inclusion of light sea quarks has been made possible 
by a formulation for light quarks on the lattice called 
improved staggered (asqtad) quarks~\cite{impstagg}. Apart from being numerically 
very fast, this formulation of the QCD action for light 
quarks is also very accurate. It has all sources of discretisation 
error removed to $\cal{O}$$(\alpha_sa^2)$, where $a$ is the 
lattice spacing and $\alpha_s$ is the QCD coupling constant. 
The gluonic piece of the action is 
also simulated with a very accurate discretisation of 
QCD which has errors at $\cal{O}$$(\alpha_s^2a^2)$ apart 
from effects from sea quarks, expected to be small, 
which feed in at $\cal{O}$$(\alpha_sa^2)$~\cite{glue}.  
In order to check the effect of discretisation errors, 
two sets of ensembles are available with 
lattice spacing values around 0.12fm and 0.09fm~\cite{milc2}, 
and recently a further, coarser set, has become available 
with a lattice spacing around 0.17fm. 

The physical volumes of the configurations are large, 
$(2.5 {\rm fm})^3$, so we do not expect significant errors from 
having a finite volume. Some analysis of finite volume errors has 
been done for light hadrons by the MILC collaboration~\cite{milc2}
and no significant effect was found on these volumes. $\Upsilon$ states 
are smaller than light hadrons in general so we 
expect even less of an effect here. 

\begin{table*}[t]
\centerline{\begin{tabular}{ccccccccc}
\hline
\hline
Lattice &$n_f$&$\beta$ &$am_l,am_s$&$u_{0L}$&$aM^0_b$ & $n_{conf}$ & $n_{orig}$\\
\hline
$16^3\times 48$ super-coarse & 2+1&6.458&0.0082,0.082 &0.814 &4.0&  461 & 8 \\
\hline
$20^3\times 64$ coarse & 0&8.0&- &0.856 &2.8&  210 & 16 \\
&2 &7.2& 0.02,- &0.845&2.8&  210 & 16 \\
&3 &6.85 &0.05,0.05&0.8391&2.8 &  210 & 8  \\
&2+1&6.81 & 0.03,0.05&0.8378&2.8 &  210 & 16 \\
&2+1&6.79 & 0.02,0.05&0.837&2.8 &  210 & 16 \\
&2+1&6.76 & 0.01,0.05&0.836&2.8 &  210 & 16 \\
\hline
$28^3\times 96$ fine&0 &8.4& -&0.8652 &1.95  & 210 & 16 \\ 
&2+1 &7.11& 0.0124, 0.031&0.855&1.95  & 210 &16  \\ 
&2+1 &7.09& 0.0062, 0.031&0.8461&1.95  & 159 & 16,12* \\ 
\hline
\hline
\end{tabular}}
\caption{Parameters and details of MILC configurations used for $\Upsilon$ correlator calculations. 
$n_f$ is the number of sea flavors, varying from 0 (quenched) 
through 2 (degenerate $u$ and $d$ only) and 3 (degenerate $u$, $d$ and $s$) to 
the most physical 2+1 (degenerate $u$ and $d$ and more massive $s$). 
$\beta$ is 
the bare gauge coupling parameter in the gluon action, $10/g^2$. 
$am_l$, $am_s$ are the light sea $u/d$ and $s$ quark masses 
respectively in lattice units 
(using the MILC convention 
for quark mass, see~\cite{ourms}). 
$u_{0L}$ is our estimate of the maximal trace link in lattice 
Landau gauge used to tadpole-improve fields in our NRQCD action. $a\mbare$ is 
the bare $b$ quark mass in lattice units. 
$n_{conf}$ is the number of 
configurations from the MILC ensemble used in our calculation 
with $n_{orig}$ origins for the heavy quark propagators per configuration.  
*For 92 of the 159 configurations, only 12 origins were used instead of 16. 
Tables~\ref{tab:ainv_superc},~\ref{tab:ainv_coarse} 
and~\ref{tab:ainv_fine} give our lattice spacing 
determinations for these ensembles; they are approximately 
0.17fm for the super-coarse ensemble, 0.12fm for the coarse 
ensembles and 0.09fm for the fine ensembles. The pion masses 
corresponding to the $u/d$ quark mass on the most chiral (i.e. lightest 
$u/d$ mass) fully unquenched 
ensemble 
in each set are 270 MeV on the super-coarse~\cite{milcprivate}, 360 MeV
on the coarse and 330 MeV on the fine ensembles~\cite{milc2}.  }
\label{tab:latdata}
\end{table*}

The sets of ensembles of gluon field configurations 
on which we have calculated bottomonium correlators 
are given in Table \ref{tab:latdata}. The ensembles within 
the `coarse' set and the `fine' set are matched to 
have approximately the same lattice spacing to separate discretisation 
effects from effects of changing the sea quark mass. 
There are results for a variety of $u/d$ quark masses, 
heavier than the real world values and we will examine 
the dependence of our results on the $u/d$ quark mass, 
aiming for conclusions that are relevant to the physical 
(chiral) limit where the $u/d$ quark mass is very small. 
The sea $s$ quark mass takes only one value here, 
which is only approximately tuned (for example, it is 
slightly high on the coarse set of ensembles). Again, 
whether this has any impact or not depends on how sensitive 
results are to the sea quark mass. 
The $b$ quarks are treated only as valence quarks because 
their effect in the sea should be negligible, being suppressed 
by inverse powers of the $b$ quark mass~\cite{nobes}. We also neglect 
$c$ quark vacuum polarization effects for the same reason. 

For the valence 
$b$ quarks in the $\Upsilon$ system we use lattice Nonrelativistic QCD (NRQCD) 
which has been developed over many years~\cite{thacker,nakhleh,upspaper} to
handle well the physics of heavy quark systems on 
the lattice. The main point here is that the $b$ quark has a mass 
which is larger than 1 in lattice units and therefore momentum 
scales of order the mass cannot be simulated without large 
discretisation errors. On the other hand such large scales do not 
need to be simulated on the lattice because they are irrelevant 
to the internal dynamics of the bound state which sets the mass 
splittings. The $b$ quark is non-relativistic inside its bound 
states ($v_b^2 \approx 0.1$ for the $\Upsilon$) and so a 
non-relativistic expansion of the QCD action is appropriate that
accurately handles scales of the order of typical momenta and 
kinetic energies inside these states. Non-relativistic QCD can 
be matched to full QCD order by order in the expansion in 
$v^2$ and $\alpha_s$. 

The lattice NRQCD Hamiltonian that we use is given by
 \begin{eqnarray}
 H &=& H_0 + \delta H; \nonumber \\
 H_0 &=& - {\delsq\over2\mbare}, \nonumber \\
\delta H
&=& - c_1 \frac{(\delsq)^2}{8(\mbare)^3}
            + c_2 \frac{ig}{8(\mbare)^2}\left(\delv\cdot\Ev -
\Ev\cdot\delv\right) \nl
 & & - c_3 \frac{g}{8(\mbare)^2} \sigmav\cdot(\delv\times\tilde{\Ev} -
\tilde{\Ev}\times\delv) \nl
 & & - c_4 \frac{g}{2\mbare}\,\sigmav\cdot\tilde{\Bv}  
  + c_5 \frac{a^2\delfour}{24\mbare} \nl
 & & -  c_6 \frac{a(\delsq)^2}{16n(\mbare)^2} .
\label{deltaH}
\end{eqnarray}
Here $\del$ is the symmetric lattice derivative and 
$\delfour$ is the lattice discretisation of the 
continuum $\sum_iD_i^4$. $\mbare$ is the bare $b$ quark mass. 
The spin-independent terms 
contribute to the (spin-averaged) differences in mass 
between radial and orbital excitations in the spectrum 
and the ground-state. 
Spin-independent terms are included through next-to-leading 
order above with the Darwin and $p^4$ relativistic corrections. 
We will now discuss remaining sources of systematic error in turn 
in the radial and orbital splittings and estimate their size. 

\begin{table*}[t]
\centerline{\begin{tabular}{ccccc}
\hline
\hline
Correction & relativistic & radiative & radiative & {\bf Total} \\
& & kinetic & Darwin & relativistic  \\
& & & & + radiative \\
\hline
Form 
& $\delta p^6/(M_b)^5 $
& $\alpha_s\delta p^4/4(M_b)^3 $
& $4\pi\alpha_s^2\psi(0)^2/(3M_b^2)$
&
\\
\hline
Est. \%age in $2S-1S$ 
& & & & 
\\
supercoarse
&0.5
&0.5
&0.5
&1.0
\\
coarse
&0.5
&0.4
&0.3
&0.7
\\
fine
&0.5
&0.3
&0.2
&0.6
\\
\hline
Est. \%age in $1P-1S$ 
& & & &
\\
supercoarse
&1.0
&2.0
&1.0
&2.5
\\
coarse
&1.0
&1.5
&0.7
&2.0
\\
fine
&1.0
&1.2
&0.4
&1.5
\\
\hline
\hline
\end{tabular}}
\caption{An estimate of systematic errors in the $2S-1S$ and $1P-1S$ 
splittings in the $\Upsilon$ in our lattice QCD calculation arising 
from missing higher order relativistic and radiative corrections 
to the NRQCD action that we use.
}
\label{tab:systrelrad}
\end{table*}
\begin{table*}[t]
\centerline{\begin{tabular}{ccccc}
\hline
\hline
Correction & discretisation in & discretisation in & discretisation in & {\bf Total} \\
& NRQCD action (i) & NRQCD action (ii) & gluon action & discretisation \\
\hline
Form 
& $\alpha_sa\delta p^4/8n(M_b)^2 $
& $\alpha_sa^2\delta p_i^4/12M_b $
& $4\pi\alpha_sa^2\psi(0)^2/15$
&
\\
\hline
Est. \%age in $2S-1S$ 
& & & & 
\\
supercoarse
&0.5
&1.7
&1.0
&2.0
\\
coarse
&0.3
&0.7
&0.5
&1.0
\\
fine
&0.2
&0.3
&0.2
&0.4
\\
\hline
Est. \%age in $1P-1S$ 
& & & &
\\
supercoarse
&2
&7
&3
&8
\\
coarse
&1.2
&3
&1.7
&4
\\
fine
&0.7
&1.2
&0.5
&1.5
\\
\hline
\hline
\end{tabular}}
\caption{An estimate of systematic errors in the $2S-1S$ and $1P-1S$ 
splittings in the $\Upsilon$ in our lattice QCD calculation arising 
from discretisation errors in the NRQCD and gluon actions. }
\label{tab:systdisc}
\end{table*}

\subsection{Systematic errors}

{\it Relativistic errors}. Errors from missing higher orders 
in the relativistic expansion 
should be $\cal{O}$$(v^4)=1\%$ in the radiative and orbital 
splittings in principle. 
However, this does not allow for the fact that the error is 
set by the {\it difference} in the expectation value of 
the relativistic correction in the two states that 
make up the splitting. 
Estimates of the 
expectation value of powers of $p$ in a potential 
model show that the $2S$ and $1S$ states are similar enough, and 
have similar enough expectation values, 
that the error in the difference between these states is significantly less 
than the naive expectation. We take the systematic error 
from relativistic corrections to be 0.5\% for the $2S-1S$ 
splitting and 1\% for the the $1P-1S$ splitting here.  
These sources of error are tabulated in 
Table~\ref{tab:systrelrad}.

{\it Radiative errors}. We must also consider missing 
radiative corrections to the 
terms that we are including in the NRQCD action. These are 
represented by the coefficients $c_i$ 
that are required to match full QCD by taking 
account of gluon radiation above 
the lattice cut-off, missing from the lattice theory. The 
$c_i$ coefficients take the form $1+c_i^{(1)}\alpha_s + \ldots$, 
where $c_i^{(1)}$ is a function of the bare heavy quark 
mass in lattice units, $M_b^0a$. 
Calculation of the $c_i^{(1)}$ is in progress 
for this action. They were calculated earlier for a similar 
but different NRQCD (and gluon) action~\cite{colin2} and found to have 
a value around 0.5 for $aM_b^0 > 1$.  
Here we have set the $c_i$ to 1, giving expected systematic 
errors of $\cal{O}$$(v^2\alpha_s)=2-3\%$ in spin-independent splittings
from a very naive analysis. 
A more sophisticated analysis, as for the relativistic corrections, 
uses the expectation 
value of $p^4$ in the different states from a potential model 
and calculates $2\alpha_s\delta(p^4)/8M_b^3$, 
where $\delta(p^4)$ is the difference in $p^4$ values 
appropriate to the splitting and the factor of 2 comes from 
having 2 $b$ quarks in an $\Upsilon$. 
The error estimates that this gives are shown in Table~\ref{tab:systrelrad}.
The effect of the Darwin 
term, $\delv\cdot\Ev$, can be estimated in a potential model 
by adding a delta function 
of appropriate strength ($\nabla^2 V$) at the origin. 
The systematic error from missing radiative corrections to the Darwin term
is then $4\pi \alpha_s^2\psi^2(0)/(3M_b^2)$ where $\psi^2(0)$ is the 
square of the `wavefunction at the origin'. We estimate this using 
our results in section~\ref{sec:lepwid} and tabulate the estimates 
in Table~\ref{tab:systrelrad}.
We can then add all our estimates together in quadrature to 
obtain a total estimated systematic error from relativistic 
and radiative corrections and this is also given in 
Table~\ref{tab:systrelrad}. The maximum error is 2.5\% 
for the $1P-1S$ splitting on the super-coarse lattices. 

Note that we 
have tadpole-improved all the gauge fields appearing above
by dividing the gluon fields, $U_{\mu}$, by a parameter 
$u_0$ which cancels, in a mean-field way, the effect of 
a disparity between the lattice and continuum gluon fields 
induced by the fact that the lattice field is exponentially 
related to the continuum field~\cite{tadimp}. Extra tadpole diagrams 
appear in the quark-gluon coupling on the lattice and these 
take a fairly universal form, allowing a cancellation by 
$u_0$. $u_0$ should be a measure of the difference between 
$U_{\mu}$ and the unit matrix, but is not uniquely specified. 
Here (as given in Table~\ref{tab:latdata}) we use $u_{0L}$, an 
estimate of the trace of the average $U_{\mu}$ field  in 
lattice Landau gauge (which maximises this trace). Subsequent 
determinations of the trace of the Landau link~\cite{milcprivate} 
have shown that 
the values used on the fine ensemble are underestimates by 
about 1\%. In Figure~\ref{fig:splitsys}, however, we show that using a rather 
different value of $u_0$ based on the plaquette ($u_{0P}=\sqrt[4]{plaq}$ = 
0.8677~\cite{milc1} on the coarse 2+1 flavor 0.01/0.05 ensemble used 
for comparison) has 
no discernible
effect (less than our 1.5\% statistical error) on spin-independent, radial
and orbital, splittings~\cite{comment}.
This also lends support 
to the idea that the $c_i$ are not very different from 1 because 
they must perturbatively correct for the different $u_0$ 
factors. Without tadpole-improvement the $c_i$ would be very 
different from 1.   

{\it Discretisation errors}. The Hamiltonian above, Equation~\ref{deltaH}, also includes 
terms which correct for discretisation errors in 
the leading order spatial derivative (squared) and temporal derivative. 
These are the terms multiplied 
by $c_5$ and $c_6$ respectively. Note that both terms have 
large numerical factors in the denominator. As above, $c_5$ 
and $c_6$ are set to 1 here, and the systematic 
error comes from neglected radiative corrections. 
The $\cal{O}$$(\alpha_s)$ pieces of 
$c_5$ and $c_6$ were calculated in ~\cite{colin2}. Indeed
$c_1$ and $c_6$ are not independent since these terms can 
be combined. Both $c_5^{(1)}$ and $c_6^{(1)}$ were found 
to be 0.5, i.e.
$\cal{O}$$(1)$, as expected. The expected systematic error 
in spin-independent splittings is then 
$\cal{O}$$(\alpha_s(a\mbare)^2v^2/12)=5\%\times(a^2 \,{\rm in \, GeV^{-2}})$
and $\cal{O}$$(\alpha_s(a\mbare)v^2/8n) < 1\%\times(a \,{\rm in \, GeV})$ from a naive 
analysis. The more sophisticated analysis as above 
uses estimates of $p^4$ in a potential model and the 
results are tabulated in Table~\ref{tab:systdisc}.
Corrections 
which are higher order in $a$ will have an even smaller effect 
because the additional powers of $a\mbare$ will come with 
additional powers of $v^2$. The Darwin term also contributes 
to the discretisation errors and an estimate of 
this is in progress. Preliminary results~\cite{nobespriv} suggest that 
it has a negligible coefficient and so we do 
not include an error for it here. 
There are, however, additional discretisation errors coming from the 
gluon action. These were previously estimated for unimproved 
glue in earlier work~\cite{oldalpha}. 
Because the error is proportional 
to the difference in the square of the wave function at the 
origin, again we find a bigger systematic error in the 
$1P-1S$ splitting than in the $2S-1S$. 
This correction is now reduced by a factor of $\alpha_s$ 
for the improved glue configurations we are working with 
here. We give estimates for its size in Table~\ref{tab:systdisc}. 
The total systematic error estimated from discretisation 
effects is also given in this Table. 
The error is sizeable on the super-coarse lattices 
and dominates that from missing relativistic and radiative 
corrections. On the fine lattices the estimated 
discretisation error is only 0.4\% in the $2S-1S$ splitting, 
about the same size as the relativistic/radiative errors.  

From our error analysis it 
is clear that we should use the $2S-1S$ splitting to set the 
lattice spacing, despite the fact that it is somewhat harder 
to obtain precise results. Then the systematic error above is smaller than 
our statistical error on all except the super-coarse ensemble. 
Because 
we have results at three different values of the lattice 
spacing we can assess how good our estimates of systematic 
errors are, and we will do that below.  

It is interesting to note that the effect of sea
$c$ and $b$ quarks is to modify the coefficients of 
the higher order improvement terms in the gluon 
action~\cite{nobes} at $\cal{O}$$(\alpha_s)$. The additional 
coefficients are very small, 0.025$\alpha_s$ for 
the rectangle term for $am_c \approx 1$, 
so from the arguments above we can see that their effect 
would be entirely negligible in spin-independent splittings. 

$E$ and $B$ are the chromoelectric and 
chromomagnetic fields. They are generally defined by standard 
cloverleaf operators (tadpole-improved) but here we use improved operators 
$\tilde{E}$ and $\tilde{B}$ for the $E$ and $B$ operators which appear in the spin-dependent 
terms. These terms are only included at leading order in 
the Hamiltonian above and therefore we expect systematic 
errors in the spin-splittings of $\cal{O}$$(v^2)=10\%$ from 
higher orders in the relativistic expansion and 
$\cal{O}$$(\alpha_s)=20\%$ from perturbative corrections 
in $c_3$ and $c_4$. Previous studies~\cite{scaling} showed that 
$a^2$ errors in the cloverleaf operators could have a substantial 
effect for the fine structure splittings, such as the 
hyperfine splitting, presumably because this splitting is 
sensitive to short distances and therefore rather large 
momenta, $pa$, for the quark. Here we therefore attempt 
to ameliorate this effect by correcting the cloverleaf 
operator in these terms for the leading discretisation 
errors. This is done by the following replacement~\cite{nakhleh} (
using the corrected tadpole-improvement factors from~\cite{clover}): 
\begin{eqnarray}
F_{\mu\nu}(x) &\rightarrow& \frac{5}{3}F_{\mu\nu}(x) 
- \frac{1}{6}[U_{\mu}(x)F_{\mu\nu}(x+a\hat{\mu})U^{\dag}_{\mu}(x) \nl
&+& U^{\dag}_{\mu}(x-a\hat{\mu})F_{\mu\nu}(x-a\hat{\mu})U_{\mu}(x-a\hat{\mu}) \nl
&-& ( \mu \Leftrightarrow \nu) ] + \frac{1}{3}(\frac{1}{u_0^2}-1)F_{\mu\nu}(x)
\label{impEB}
\end{eqnarray}
where all gluon fields are understood to be tadpole-improved. 

\subsection{Smearing and Fitting}

The $b$ quark propagators (Green's functions) are calculated on one pass through 
the lattice, using the evolution equation:
\begin{eqnarray}
G(\vec{x},t+1) &=& (1-\frac{a\delta H}{2})(1-\frac{aH_0}{2n})^nU^{\dag}_{t}(x) \nl
        & & (1-\frac{aH_0}{2n})^n(1-\frac{a\delta H}{2}) G(\vec{x},t) 
\end{eqnarray}
with starting condition:
\begin{equation}
G(\vec{x},0) = \phi(x).
\end{equation}
$\phi(x)$ is a smearing function which, when the quark propagators 
are combined into meson correlators, improves the overlap 
with particular ground and excited states for a better signal. 
This will be discussed further below. $n$ is a stability parameter 
which is chosen to tame (unphysical) high momentum modes of 
the $b$ quark propagator which might otherwise cause the meson 
correlators to grow exponentially with time rather than 
fall. It is numerically more convenient and safer to fit an exponentially 
falling correlator.

\begin{figure}[t]
\includegraphics[width=7cm,angle=-90]{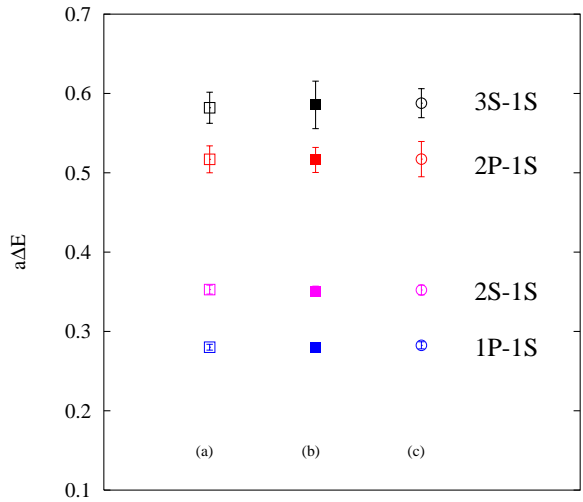}
\caption{Dependence of $\Upsilon$ energy splittings on change in NRQCD stability parameter $n$ and choice of $u_0$: (a) $n=2$, $u_{0L}$ (b) $n=4$, $u_{0L}$ (c) $n=2$, $u_{0P}$. The results are from the 0.01/0.05 2+1 flavor coarse ensemble. $2S-1S$ etc indicates the 
splitting between radial excitations of the $\Upsilon$ and $1P-1S$ the 
splitting between the appropriate $^1P_1$ state and the $\Upsilon$. }\label{fig:splitsys}
\end{figure}

The calculations presented in this paper incorporate a NRQCD stability parameter 
value of $n=2$. Physical results should be independent of $n$ for reasonable values 
of $n$ and we checked that by repeating
calculations on the coarse 2+1 flavor $m_{u/d}/m_s=0.01/0.05$ ensemble 
with $n=4$. Figure \ref{fig:splitsys} shows that no dependence on $n$ was found in the 
resulting $\Upsilon$ energy splittings. 

During tuning runs to fix the $b$ quark mass we obtained 
results with lower statistics for several 
values of $M_b^0$ varying by $\cal{O}$(10\%). We found no 
discernible difference in $1P-1S$ or $2S-1S$ splittings within 
our statistical/fitting errors. This is an expected feature
of radial and orbital splittings in the heavyonium sector, 
since these splittings are observed experimentally to vary 
very little between charmonium and bottomonium. 
It is one of the reasons why these splittings are useful 
to set the lattice spacing for an ensemble of configurations, 
because it can be done reliably without a precise tuning 
of the $b$ quark mass. 

We use 4 different smearing functions for $S$-wave states. One 
is a local smearing or delta function, $\phi_{loc}(x)=\delta_{x,0}$ where 
0 is the source point. The other 3 smearings are set up 
as simple functions of radial distance $r$ from the source point. 
This requires that the gluon fields are fixed to Coulomb gauge. 
Then a wavefunction picture makes sense and we do not have to
insert any gauge links into the smearing function. 
The smearings are labelled:
\begin{eqnarray}
\phi_{gs}(r) &=& \exp(-\frac{r}{a_0}) \nonumber \\
\phi_{es}(r) &=& (2a_0-r)\exp(-\frac{r}{2a_0}) \nonumber \\
\phi_{ds}(r) &=& (27a_0^2-18a_0r+2r^2)\exp(-\frac{r}{3a_0}) \nonumber \\
\end{eqnarray}
for good overlap with the ground, first excited and second 
(doubly) excited $\Upsilon$ states.  We found $a_0$=1.0 in 
lattice units to be a good value on testing the effect 
of smearing and have used this value on 
all three sets of ensembles, despite the change in lattice 
spacing. 
In fact the $\phi_d$ smearing gives results very similar to the 
$\phi_e$ smearing and needs to be improved to be more useful. 
For $P$-wave states we use $g$ and $e$ smearing functions also based 
on hydrogen-like wavefunctions, i.e. for the $g$ smearing 
we take $x, y$ and $z$ times $exp(-{r}/{2a_0})$ in each 
of the 3 coordinate directions. For the $e$ smearing for 
$P$-waves we take
$r$ dependence of the form $(6a_0-r)exp(-{r}/{3a_0})$, 
again multiplied by $x$, $y$ or $z$ for each direction.
We find that it 
is better to take $a_0$ = 0.5 for the $P$-wave states. 
For $D$-wave states for the $g$ smearing function 
we have used the same $r$ dependence and $a_0$ value as 
for the $P$-wave states, but multiplied by $xy$, $yz$ or $zx$. 
Instead of an $e$ type smearing we simply used sources made 
of an appropriate set of delta functions, 1 lattice spacing 
from the origin. 

We use fits to a matrix of correlators with different smearings at
source and sink of the form:
\begin{equation}
G_{\mathrm{meson}}(n_{sc},n_{sk};t) = \sum_{k=1}^{n_{exp}}a(n_{sc},k)a^*(n_{sk},k)e^{-E_kt}.\label{eq:fit}
\end{equation}
$a(n_{sc/sk},k)$ are the (real) amplitudes for state $k$ to appear in the 
smearings used at source and sink respectively. 
We use a constrained fitting method~\cite{bayes} which allows a large number 
of exponentials to be used in the fit by constraining the way 
in which these exponentials can appear based on physical 
information. 

The augmentation of $\chi^2$ with a Bayesian term, 
\beq
\chi^2\rightarrow\chi^2_\mathrm{aug}\equiv\chi^2+\chi^2_\mathrm{prior}
\label{}
\eeq
where
\beq
\chi^2_\mathrm{prior}\equiv\sum_{k}\frac{(p_k-\tilde{p}_k)^2}{\tilde{\sigma}_{p_k}^2}
\label{eq:bayesdef}
\eeq
allows the data over the entire range of $t$ to be fitted to 
any number of exponentials in Equation~\ref{eq:fit}.
The fit parameters, $p_k$, were taken as the amplitudes $A_n$, 
the log of the ground state energy $\ln(E_0)$ and the logs of the 
energy differences $\ln(E_{n+1}-E_n)$. 
This choice prevents the fit from venturing into unphysical 
negative energy space and ensures the correct ordering of states. 
Initial fits were performed with tight widths $\tilde{\sigma}_{p_k}$ 
on the lowest energy parameters. The resulting fit parameters were 
then used as the prior $\tilde{p}_k$ values in the final fits where all 
widths were relaxed to $\tilde{\sigma}_{p_k}=1$.

\begin{figure}[t]
\includegraphics[width=7cm,angle=-90]{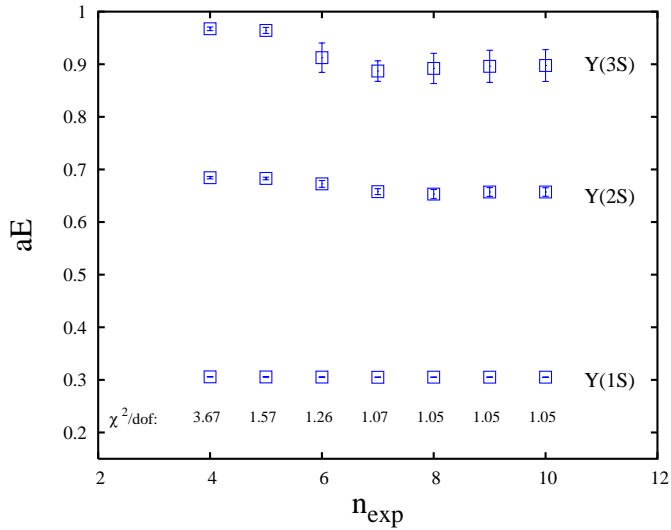}
\caption{$\Upsilon$ energies in lattice units as a function of number of exponentials used in a $3\times 3$ matrix fit for results on the 0.01/0.05 coarse ensemble. 
$\chi^2/\mbox{dof}$ values indicate goodness of fit.}\label{fig:EvsNexp}
\end{figure}

Figure \ref{fig:EvsNexp} shows the dependence of the resulting energies 
on the number of exponentials used in an example $3\times 3$ matrix fit 
(to correlators from the 0.01/0.05 2+1 flavor ensemble). 
Notice that the fit results become independent of $n_{exp}$, both 
in terms of fit error and in terms of goodness of fit for 
$n_{exp} > 6$. 
This is the usefulness of the Bayesian approach.
It is very simple to choose a fit result because there is 
no trade-off between worries of systematic contamination from 
higher order exponentials and the fit error. Any result with $n_{exp}>6$ 
will give the same result. 
In this case $n_{exp}=7$ was chosen. 
Notice that
the energies of the higher radial excitations 
require a larger $n_{exp}$ to converge, reasonably enough.
If the only interest here was in the ground state energy fits 
from a smaller $n_{exp}$ value would be as good. 

\section{Results}\label{sec:results}
\begin{table}[t]
\centerline{\begin{tabular}{cc}
\hline
\hline
& $0.0082/0.082$ \\
\hline
$aE(1^3S_1)$&0.27755(36) \\
$aE(2^3S_1)$&0.770(8) \\
$aE(1^1P_1)$&0.6674(64) \\
\hline
\hline
\end{tabular}}
\caption{Fitted energies in lattice units for various radial 
and orbital excitations on the super-coarse ensemble.}
\label{tab:SI_E_superc}
\end{table}
\begin{table*}[t]
\centerline{\begin{tabular}{ccccccc}
\hline
\hline
& $0.01/0.05$ & $0.02/0.05$ & $0.03/0.05$ & $0.05/0.05$ & 0.02 &Quenched\\
\hline
$aE(1^3S_1)$&0.30506(27)
&0.30118(25)
&0.29848(26)
&0.29588(34)
&0.29563(26)
&0.26621(24)\\
$aE(2^3S_1)$&0.6578(56)
&0.6519(44)
&0.627(10)
&0.6381(95)
&0.6326(79)
&0.6073(49) \\
$aE(3^3S_1)$&0.887(20)
&0.878(22)
&0.870(32)
& 0.927(21)
&0.851(33)
& 0.887(12)\\
$aE(1^1P_1)$&0.5852(37)
&0.5705(41)
&0.5593(70)
&0.5593(61)
&0.5519(59)
&0.5111(27)\\ 
$aE(2^1P_1)$&0.822(17)
&0.801(16)
&0.784(53)
&0.827(18)
&0.787(29)
&0.788(12) \\
$aE(1^3D_2)$&0.751(12)
&-
&-
&-
&-
&-\\
$aE(2^3D_2)$&1.026(34)
&-
&-
&-
&-
&- \\
\hline
\hline
\end{tabular}}
\caption{Fitted energies in lattice units for different radial 
and orbital excitations for the coarse lattice ensembles. }
\label{tab:SI_E_coarse}
\end{table*}
\begin{table}[t]
\centerline{\begin{tabular}{cccc}
\hline
\hline
& $0.0062/0.031$ & $0.0124/0.031$ & Quenched\\
\hline
$aE(1^3S_1)$&0.23248(46)
&0.28677(40)
&0.18517(23)\\
$aE(2^3S_1)$&0.4818(33)
&0.5303(27)
&0.42734(25)\\
$aE(3^3S_1)$&0.619(14)
&0.671(13)
&0.6472(71)\\
$aE(1^1P_1)$&0.4233(37)
&0.4705(30)
&0.3595(19)\\
$aE(2^1P_1)$&0.574(44)
&0.628(16)
&0.5470(86)\\
\hline
\hline
\end{tabular}}
\caption{Fitted energies in lattice units for various radial and orbital excitations on the fine lattice ensembles.}
\label{tab:SI_E_fine}
\end{table}
\begin{table}[t]
\centerline{\begin{tabular}{ccc}
\hline
\hline
& $n=4$ & $u_{0}=\sqrt[4]{plaq}$ \\
\hline
$aE(1^3S_1)$&0.29433(27)
&0.47736(25)\\
$aE(2^3S_1)$&0.6449(65)
&0.8296(62)\\
$aE(3^3S_1)$&0.8799(65)
&1.0652(62)\\
$aE(1^1P_1)$&0.5743(37)
&0.7599(45)\\
$aE(2^1P_1)$&0.811(16)
&0.995(22)\\
\hline
\hline
\end{tabular}}
\caption{$\Upsilon$ spin independent energy splittings in lattice units on the 
coarse 2+1 flavor 0.01/0.05 ensemble. These were used for checking 
systematic errors by comparison with results in Table~\ref{tab:SI_E_coarse}.}
\label{tab:SI_sys}
\end{table}

$3\times 3$ or $4\times 4$ matrix fits were performed 
to the $S$-wave correlators, 
and $2\times 2$ fits performed to the $P$- and $D$-wave  correlators. 
$D$-wave correlators have only been calculated on the most chiral (0.01/0.05) 
2+1 flavor coarse 
ensemble, and only for a subset of $D$-wave spin states. 
For each fit, the best $n_{exp}$ was chosen as one which gave 
a reasonable  $\chi^2/\mbox{dof}$ and was large enough for fitted results to 
be independent of $n_{exp}$. 
The results for the fit energies are given in Tables \ref{tab:SI_E_superc}, \ref{tab:SI_E_coarse} 
and \ref{tab:SI_E_fine} for the super-coarse, coarse and 
fine lattices respectively. It is 
clear that the fitted energies are very accurate when compared to 
previous results, even from the quenched approximation~\cite{scaling}. This is 
partly because of the large number of configurations, their large 
physical volume and the large number of origins that 
we have been able to use. It is also partly because of the improved fitting 
strategy. 

In the next two sections we will discuss the spin-independent aspects 
of the spectrum, that is the differences in mass between 
different radial and orbital excitations. In principle we should 
be using masses to calculate the mass splittings that are 
spin-averaged over the different spin states for each excitation. 
However, we cannot do that for the $S$-wave states since no 
$^1S_0$ state has been seen experimentally. For the $P$-wave states 
we can spin-average but it is easier to take simply the 
mass of the $^1P_1$ state. In potential 
models, where the spin-spin coupling induces a delta-function 
potential, the difference between the $^1P_1$ mass and that 
of the spin-average of the $^3P_{0,1,2}$ is expected to be zero. 
We will examine the limits to that assumption in section~\ref{sec:fine}.
In the meantime we will work with the $^3S_1$ and $^1P_1$ results 
from the Tables.

For completeness we also give the results for different values of 
stability parameter $n$ and tadpole-improvement factor $u_0$ 
for the 0.01/0.05 coarse ensemble in Table~\ref{tab:SI_sys}. 
It is clear from this table that, although changing $u_0$ does not affect 
spin-independent splittings, it does provide an overall shift to the  
energies.

\section{Determining the lattice scale}\label{sec:scale}
\begin{table}[t]
\centerline{\begin{tabular}{cc}
\hline
\hline
& $0.0082/0.082$ \\
\hline
$a^{-1}_{2S-1S}$(GeV)
&1.144(19)(11)(23)
\\
$a^{-1}_{1P-1S}$(GeV)
&1.128(19)(28)(90)
\\
\hline
\hline
\end{tabular}}
\caption{Inverse lattice spacing determinations for the super-coarse ensembles 
from both the $1^1P_1-1^3S_1$ and $2^3S_1-1^3S_1$ splittings. 
The first error given is statistical/fitting. The second is 
the
expected systematic error from relativistic/radiative 
corrections and the third from discretisation errors.} 
\label{tab:ainv_superc}
\end{table}
\begin{table*}[t]
\centerline{\begin{tabular}{ccccccc}
\hline
\hline
& $0.01/0.05$ & $0.02/0.05$ & $0.03/0.05$ & $0.05/0.05$ & 0.02 &Quenched\\
\hline
$a^{-1}_{2S-1S}$(GeV)
&1.596(25)(11)(16)
&1.605(20)(11)(16)
&1.714(52)(12)(17)
&1.645(46)(12)(16)
&1.671(39)(12)(17)
&1.651(24)(12)(17)
\\
$a^{-1}_{1P-1S}$(GeV)
&1.571(21)(31)(63)
&1.634(25)(33)(65)
&1.687(45)(34)(67)
&1.670(39)(33)(67)
&1.717(40)(34)(68)
&1.797(20)(36)(72)
\\
\hline
\hline
\end{tabular}}
\caption{Inverse lattice spacing determinations for the coarse ensembles from both the $1^1P_1-1^3S_1$ and $2^3S_1-1^3S_1$ splittings. 
The first error given is statistical/fitting. The second is the 
expected systematic error from relativistic/radiative corrections 
and the third from discretisation errors. }
\label{tab:ainv_coarse}
\end{table*}
\begin{table*}[t]
\centerline{\begin{tabular}{ccccccc}
\hline
\hline
& $0.0062/0.031$ & $0.0124/0.031$ &Quenched\\
\hline
$a^{-1}_{2S-1S}$(GeV)
&2.258(30)(14)(9)
&2.312(26)(14)(9)
&2.324(24)(14)(9)
\\
$a^{-1}_{1P-1S}$(GeV)
&2.305(45)(35)(35)
&2.390(38)(36)(36)
&2.524(28)(38)(38)
\\
\hline
\hline
\end{tabular}}
\caption{Inverse lattice spacing determinations for the fine ensembles 
from both the $1^1P_1-1^3S_1$ and $2^3S_1-1^3S_1$ splittings. 
The first error given is statistical/fitting. The second is the
expected systematic error from relativistic/radiative corrections 
and the third from discretisation errors. }
\label{tab:ainv_fine}
\end{table*}
The lattice spacing is fixed by the choice of $\beta$ but is not determined 
until after the simulation when a physical quantity is matched to experiment. 
Once the lattice spacing has been set in this way it can be used to give 
dimensionful predictions for other observables.

There should be only one value for the lattice spacing and therefore 
it should not matter which hadron mass or mass splitting is used in fixing it. 
However, it has been a long known problem that in fact this is 
not the case in lattice simulations with unrealistic (or completely missing) 
light-quark vacuum polarization. 
This is because different physical quantities 
probe different energy scales and the strong coupling constant will 
not run correctly between these scales unless the sea quark 
content (i.e. the quark vacuum polarization contribution) 
is sufficiently correct. This gives ambiguities in calculating 
physical observables as it is impossible to say what the `right' 
lattice spacing is.  

Determinations of the lattice spacing from different splittings can then be 
used as a guide to judge how physical is the sea quark content of the 
configurations being used. Here, the orbital $1P-1S$ 
(i.e. $1^1P_1- 1^3S_1$)  and 
radial $2S-1S$ (i.e. $2^3S_1-1^3S_1$) energy splittings independently 
fix the lattice spacing~\cite{pdg}. The results are given in Tables~\ref{tab:ainv_superc}, \ref{tab:ainv_coarse} 
and \ref{tab:ainv_fine} for the super-coarse, coarse and 
fine lattices respectively. There we list the statistical/fitting uncertainties
in $a^{-1}$ and also give values for the expected systematic errors 
as estimated in Section~\ref{sec:calc}. The systematic errors 
will affect all the estimates of $a^{-1}$ in the same way, so 
it is only the differences between these errors that 
affect any discussion of differences between $a^{-1}$ 
determinations. We have not included an error from our inability 
to spin-average the $S$-wave states. The error from that 
is set by one quarter of the error in the spin-splitting 
(since we are using the vector state instead of the spin-average)
and the spin-splitting is already only 10\% of a typical 
spin-independent splitting. A 20\% error in the spin-splitting 
from missing relativistic corrections would then lead to 
a 0.5\% error in the determination of $a^{-1}$ from, for 
example, the $1P-1S$ splitting. This is negligible.

It is quite clear from these tables that there is a problem of 
internal consistency in the quenched approximation. The ratio of $a^{-1}$ 
from $1P-1S$ to that from $2S-1S$ is 1.09(2)(4) on the coarse 
quenched ensemble and 1.09(2)(1) on the fine quenched ensemble. 
The first error here is from statistics/fitting and the second 
from an estimate of the difference in systematic errors for 
the $1P-1S$ and $2S-1S$ splittings.
Previous results from quenched NRQCD calculations 
also gave ratios larger than 1~\cite{further, scaling, sesam, manke}, 
although generally with lower 
significance because of poorer statistics or fitting. 
One problem with previous results was that most of them 
used an unimproved gluon action 
and the effect of the tree-level $\cal{O}$$(a^2)$ discretisation errors 
is to reduce 
this ratio of $a^{-1}$ determinations by a few percent, 
obscuring the disagreement with 1. The ratio including a correction 
for this is given in reference~\cite{scaling} as 1.06(4) at 
$\beta$ = 6.0. Without correcting for it, reference~\cite{sesam} 
gives 1.12(9). It is clear that our new quenched results on 
improved glue are a considerable improvement in clarity on these 
previous results. 

\begin{figure}[t]
\includegraphics[width=7cm,angle=-90]{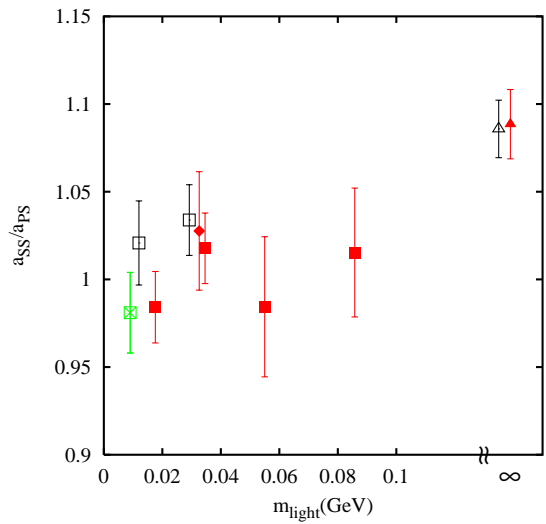}
\caption{The ratio of lattice spacing values obtained from 
the $2S-1S$ and $1P-1S$ splittings in the $\Upsilon$ system 
as a function of bare light quark mass in the lattice QCD calculation. 
Notice the very magnified $y$-axis scale.
The crossed square is from the super-coarse ensemble, the filled squares from 
the coarse 2+1 flavor ensembles and the open squares 
from the fine 2+1 flavor ensembles. The filled diamond
is from the coarse $n_f=2$ ensemble. The open and filled 
triangles on the right of the plot are results from the 
quenched fine and coarse ensembles respectively. Errors are 
statistical only. }\label{fig:ainv}
\end{figure}

Figure~\ref{fig:ainv} shows the ratio of lattice spacing 
determinations from the $2S-1S$ and $1P-1S$ splittings as a 
function of the bare light $(u/d)$ quark mass in our
simulations 
(see Table~\ref{tab:latdata}). Also included are results 
with 2 flavors of sea quarks only, and over to the right 
of the plot, the quenched results. The ratio of inverse lattice 
spacings on the most chiral 2+1 flavor ensemble at each lattice spacing 
gives 0.981(23)(60) for super-coarse, 0.984(20)(34) for coarse and 
1.021(23)(13) for fine, where the first error is statistical 
and the second systematic. All of these results are in agreement with 
1, unlike the case in the quenched approximation. The 
results agree well with each other too, and this provides a 
check on our estimates of systematic errors from discretisation 
effects. For the fine lattices discretisation effects are 
small, but on the super-coarse lattices we have estimated that 
the $1P-1S$ splitting could be moved by 6\% more than the 
$2S-1S$ splitting by discretisation and relativistic/radiative
systematic errors. This looks like an overestimate since 
our super-coarse result agrees with 1 within statistical 
errors of 2\%. 

The agreement between the inverse lattice spacing results 
from the two different splittings is confirmation of 
the approach to the physical real world of a single 
inverse lattice spacing as the sea quark content is 
made realistic. Some signs of this were seen in earlier 
2 flavor simulations~\cite{upspaper, further, sesam, marcantonio}. Indeed 
here our 2 flavor results also give agreement with 1 for the 
ratio of lattice spacing values within 
the statistical error. As discussed earlier, the lattice 
spacing from the $2S-1S$ splitting is to be preferred on 
the grounds of systematic error and so we will 
use that in what follows.  

\begin{figure}[t]
\includegraphics[width=7cm,angle=-90]{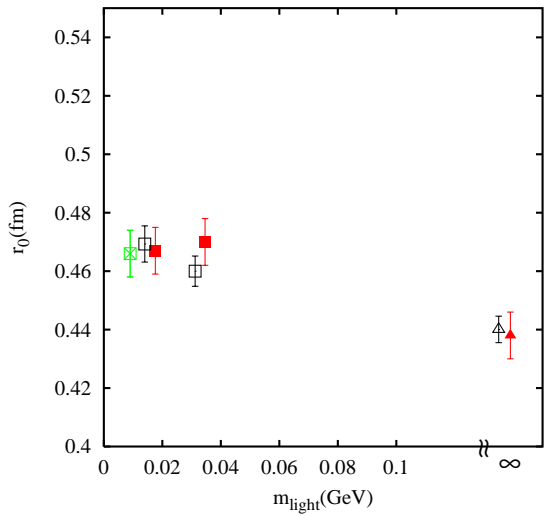}
\caption{The value of $r_0$ in fm obtained on different super-coarse, coarse and 
fine ensembles, using the $2S-1S$ splitting in the $\Upsilon$ system 
to fix the lattice spacing. 
The crossed square is from the super-coarse ensemble, the filled squares from 
the coarse 2+1 flavor ensembles and the open squares 
from the fine 2+1 flavor ensembles. 
The open and filled 
triangles on the right of the plot are results from the 
quenched fine and coarse ensembles respectively. 
The $y$-axis scale is expanded to make error 
bars visible. $r_0$ values in lattice units are 
taken from~\cite{milc1,milc2,milcprivate}. Errors include 
the statistical errors from determining the lattice spacing and from determining 
$r_0$. }\label{fig:r0}
\end{figure}

\begin{figure}[t]
\includegraphics[width=7cm,angle=-90]{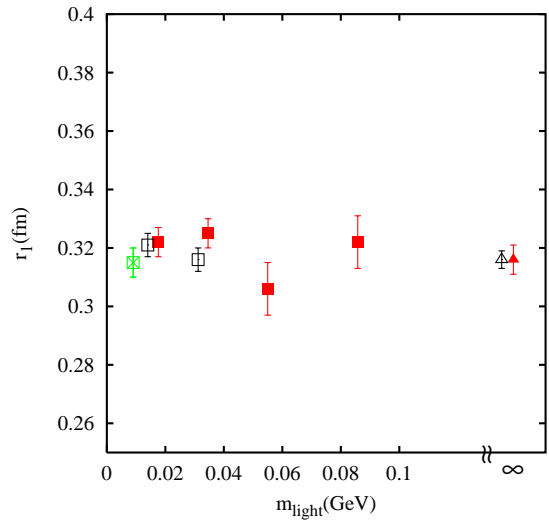}
\caption{The value of $r_1$ in fm obtained on different super-coarse, coarse and 
fine ensembles, using the $2S-1S$ splitting in the $\Upsilon$ system 
to fix the lattice spacing. 
The crossed square is from the super-coarse ensemble, the filled squares from 
the coarse 2+1 flavor ensembles and the open squares 
from the fine 2+1 flavor ensembles. 
The open and filled 
triangles on the right of the plot are results from the 
quenched fine and coarse ensembles respectively. 
The $y$-axis scale is expanded to make error bars 
visible. $r_1$ values in lattice units are 
taken from~\cite{milc1,milc2,milcprivate}. Errors include 
the statistical errors from determining the lattice spacing and from determining 
$r_1$. }\label{fig:r1}
\end{figure}

We can also ask what the lattice spacing derived from $\Upsilon$ 
splittings implies for other quantities. A quantity
that is used frequently to set the scale is the $r_0$ 
parameter~\cite{sommer}. 
It is derived from the heavy quark potential
as the value of $r$ at which $r^2F(r)=1.65$, where $F(r)$ is
the gradient of the potential. $r_0$ 
is easily and precisely calculable in lattice calculations but 
has the disadvantage that its value in physical units is 
not directly given by experiment. 
Instead its value must be given by numerical simulation on the 
lattice and comparison to an experimentally accessible quantity. 
Here we can give a value for $r_0$ from comparison to 
$\Upsilon$ splittings. 
In Figure~\ref{fig:r0} we plot $r_0$ in fm as a function of 
light sea quark mass, fixing the lattice spacing from 
the $2S-1S$ splitting. The value of $r_0$ in lattice units 
has been calculated by the MILC collaboration~\cite{milc1,milc2}. 
The result does not show any dependence on either the light 
quark mass for light enough light quark mass or the lattice spacing. 
Our best result is from the fine lattices, and gives a 
physical value for $r_0$ equal to 0.469(7)fm. 
The error includes the systematic errors in the determination 
of the lattice spacing on the fine lattices and a similar systematic 
error for the determination of $r_0$. However these errors 
are small compared to the statistical error from the 
determination of the lattice spacing. This result is 
in agreement with the results from~\cite{milc2} which were based 
on our preliminary analysis of the $\Upsilon$ results 
presented in this paper. 
We similarly have a value for the heavy quark potential variable $r_1$ ~\cite{milc2}
(where $r^2F(r)=1.00$)
of 0.321(5)fm. This is also shown 
as a function of light quark mass in Figure~\ref{fig:r1} and 
also agrees with the preliminary value from~\cite{milc2}. 
This variable in fact shows a very flat 
curve as a function of light quark mass when the scale is 
set by the $\Upsilon$ $2S-1S$ splitting, 
i.e. the light quark vacuum polarisation 
effects are very similar in both variables. Note that if we had 
used the $1P-1S$ splitting to set the lattice spacing, both 
$r_1$ and $r_0$ would have had values in fm differing by 10\% in 
the quenched approximation, because of the ambiguity in setting 
the lattice spacing demonstrated in Figure~\ref{fig:ainv}. 
The result on the chiral 2+1 flavor
configurations would be unchanged, because here there is 
no ambiguity about the value of $a$~\cite{us}.


\section{Comparing radial and orbital splittings to experiment}\label{sec:radorb}
\begin{figure}[t]
\includegraphics[width=7cm,angle=-90]{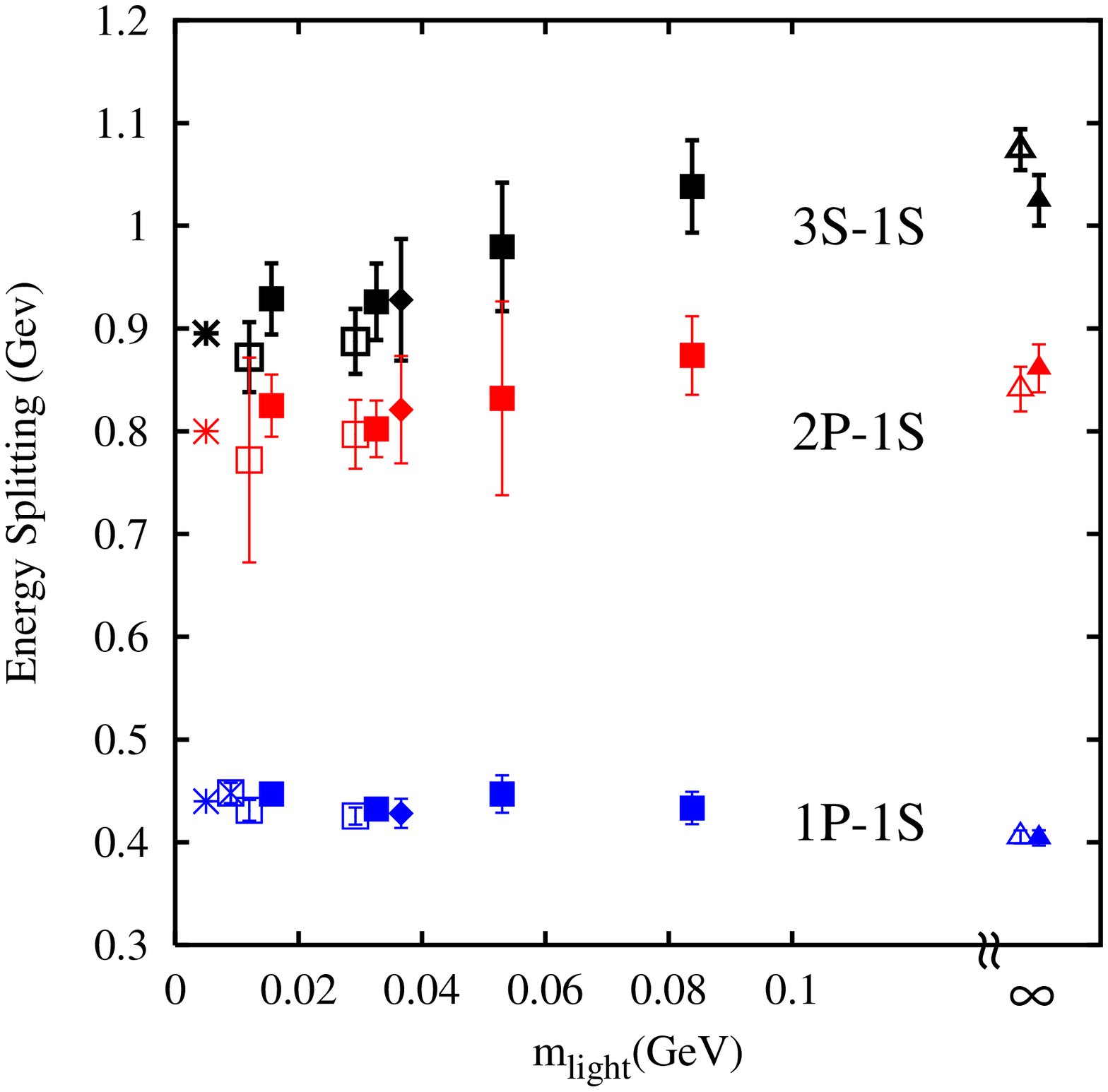}
\caption{Spin-independent splittings in the $\Upsilon$ spectrum as a function 
of bare sea light quark mass. 
The $1P-1S$, $2P-1S$, and $3S-1S$ splittings are shown.  
The $S$ state is always the $^3S_1$. The $P$ state 
from the lattice calculations is the $^1P_1$; from experiment it 
is the spin-average of $^3P_{0,1,2}$. Closed squares are from coarse 
2+1 flavor lattices and open squares from fine 2+1 flavor lattices. The crossed 
square for the $1P-1S$ splitting is from the super-coarse ensemble.  
Closed diamonds are from the coarse $n_f=2$ run. Closed and open 
triangles are from quenched coarse and fine ensembles respectively. 
Bursts give the 
experimental results. 
Errors include the statistical errors from determining 
the lattice spacing from the $2S-1S$ splitting. 
}\label{fig:split}
\end{figure}

The $2S-1S$ inverse lattice spacing values and fit energies from the previous 
sections were combined to give dimensionful determinations of the radial 
and orbital splittings
$3^3S_1-1^3S_1$, $1^1P_1-1^3S_1$, $2^1P_1-1^3S_1$.
Figure \ref{fig:split} plots the results as a function of 
the sea light quark mass for the coarse and fine
ensembles. Bursts denote the experimental values where we 
have used the spin-average of the $\chi_b$ states in 
place of the unobserved $h_b$. 

Good agreement with experiment is seen as the chiral limit is approached 
and the flatness of the approach indicates that chiral extrapolations 
would have no significant effect in shifting the results from the most chiral ensemble.

\begin{figure}[t]
\includegraphics[width=7cm,angle=-90]{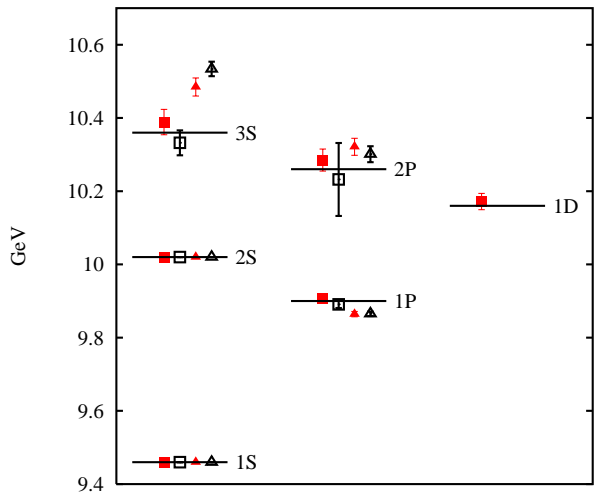}
\caption{The $\Upsilon$ spectrum of radial and orbital levels, taking 
the most chiral points from Figure~\ref{fig:split}. Closed 
and open symbols are from coarse and fine lattices respectively. 
Squares and triangles denote unquenched and quenched results respectively. 
Lines represent experiment. }\label{fig:ups_si}
\end{figure}

Figure \ref{fig:ups_si} shows the radial and orbital levels 
in the $\Upsilon$ spectrum with the lattice results from the 
coarse and fine 2+1 flavor ensembles with lightest sea
quark mass contrasted to coarse and fine quenched results.  
The solid lines are the experimental values, including the recently 
observed $^3D_2$ state~\cite{cleo}. On the lattice the spin 2 
representation is split into two representations labelled 
$E$ and $T$. Here we used the $E$ representation. This 
will be discussed further in section~\ref{sec:fine}. 
The $1S$ energy and $2S-1S$ splitting 
are not predictions as they have already been used to fix the $b$ quark mass 
in lattice units and the lattice spacing respectively. 
The quenched results clearly do not agree with experiment 
for other splittings 
whereas the 2+1 flavor results do. 

\section{The $b$ quark mass}\label{sec:mb}

The bare $b$ quark mass in the lattice Lagrangian is fixed 
by the requirement to get the $\Upsilon$ mass correct. 
Because the quark mass term has been removed from the NRQCD Lagrangian 
the energies calculated for hadrons at zero momentum are 
shifted from their masses. This means that energy differences 
(for particles containing the same number of quarks 
or antiquarks)
do give mass differences 
but in order to fix one mass absolutely, a dispersion 
relation must be calculated. In sections~\ref{sec:scale} and~\ref{sec:radorb} we 
discussed energy differences and mass splittings. Here we give results for 
the mass of the $\Upsilon$ 
from calculations of its energy as a function of lattice 
momentum, and discuss how well the $b$ quark mass 
can then be determined.  

\subsection{Fixing the $\Upsilon$ mass}

The continuum relationship between 
energy, $E$, and spatial momentum, $\vec{p}$, for a hadron in 
a theory such as NRQCD where the zero of energy has 
been shifted is given by: 
\begin{equation}
E(p) = E(0) + \sqrt{p^2+M^2} - M.
\label{eq:disp}
\end{equation}
The hadron mass is denoted by $M$ here, 
sometimes called the kinetic mass to 
emphasise that it is determined from a dispersion 
relation rather than $E(0)$. 
Equation~\ref{eq:disp} is a fully relativistic formula. 
When used in our lattice calculations, 
$\vec{p}$ is a lattice 
momentum given in lattice units by $2\pi (n_x,n_y,n_z)/L$ where 
$L$ is the number of lattice spacings in the
spatial directions (16 for super-coarse, 20 for coarse 
and 28 for fine) and $n_{x,y,z}$ take integer values between 0 and $L-1$. 
$M_{kin}$ is obtained by first fitting the energy of 
jack-knifed ratios of $\Upsilon$ correlators at 
zero and non-zero values of $p$ to an exponential form. The 
exponent yields the energy 
difference 
$\Delta E = E(p)-E(0)$ and 
\begin{equation}
M = \frac{p^2-(\Delta E)^2}{2\Delta E}.\label{eq:mkin}
\end{equation}
Note that a non-relativistic formula would not 
include the $(\Delta E)^2$ term, and this 
becomes increasingly negligible as the mass increases. 

\begin{table}[t]
\centerline{\begin{tabular}{cccc}
\hline
\hline
$n^2$&$a\Delta E$&$aM_{\Upsilon}$&$c^2$\\
\hline
1&0.008264(40)&5.967(29)&-\\
2&0.016507(84)&5.971(30)&0.999(5)\\
3&0.02473(13)&5.974(32)&0.999(5)\\
4&0.03286(19)&5.990(34)&0.996(6)\\
5&0.04104(25)&5.991(36)&0.996(6)\\
6&0.04928(31)&5.983(37)&0.997(6)\\
8&0.06540(47)&6.004(43)&0.994(7)\\
9&0.07353(57)&6.003(46)&0.994(8)\\
12&0.09778(92)&6.007(56)&0.993(9)\\
\hline
\hline
\end{tabular}}
\caption{Energy differences of zero and non-zero momentum $\Upsilon$ mesons, 
and resulting $\Upsilon$ masses, in lattice units. Non-zero momenta are 
denoted by $n^2 = n_x^2 + n_y^2 + n_z^2$. These are unambiguous except 
for $n^2=9$ which corresponds to the $n$-vector (2,2,1) and its 
permutations. The fourth column gives values for the speed of light, 
$c^2$, defined in the text. The results are for the coarse 2+1 flavor 
0.01/0.05 ensemble. }
\label{tab:mkin}
\end{table}

\begin{figure}[t]
{\includegraphics[width=7cm,angle=-90]{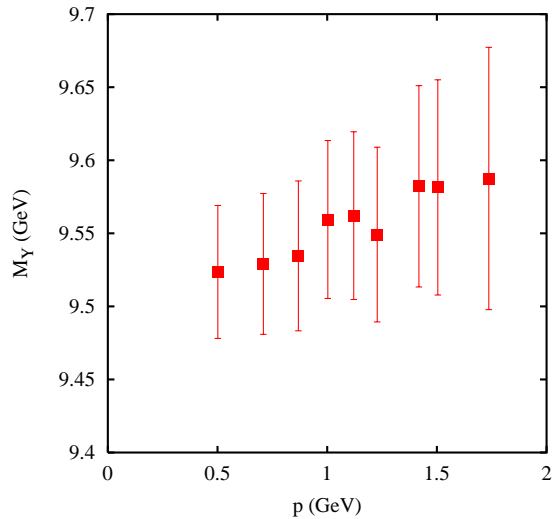}}
\caption{The value of the $\Upsilon$ mass in 
physical units plotted against the 
the lattice momentum for the $n_f=2+1$ 0.01/0.05 coarse 
ensemble for $aM^0_b=2.8$. Errors included are only those 
from determining $M_{\Upsilon}$, since the errors from determining 
the lattice spacing are the same for all points. }\label{fig:kinmass}\end{figure}

Table~\ref{tab:mkin} gives, as an example of our 
results, values for $\Delta E$ and $M$ obtained 
on the coarse $2+1$ flavor 0.01/0.05 ensemble for a range of lattice 
momenta. These results are plotted in physical units 
(using the lattice spacing from the $2S-1S$ splitting 
of section~\ref{sec:scale}) in Figure~\ref{fig:kinmass}.
Figure~\ref{fig:kinmass} shows that the $\Upsilon$
mass is stable and well-determined out to very 
large values of $p^2$, far larger than are routinely 
accessible in light hadron calculations. The value does not change 
significantly, showing 
that lattice artefact terms and errors coming from missing relativistic 
corrections to the NRQCD action in the lattice 
dispersion relation are not significant. 

\begin{figure}[t]
{\includegraphics[width=7cm,angle=-90]{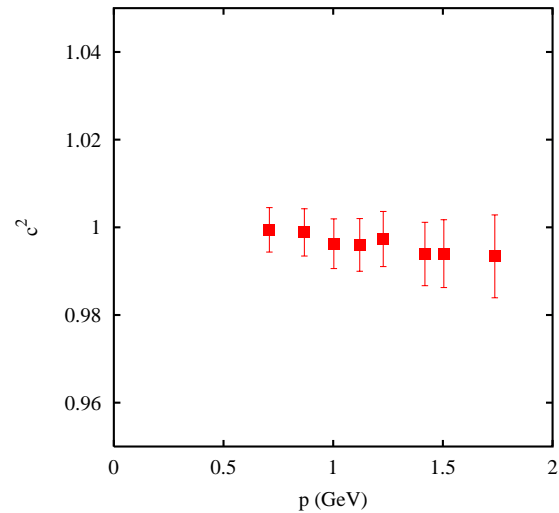}}
\caption{The `square of the speed of light' (see text 
for definition) from $\Upsilon$ correlators
at non-zero momentum 
for the $n_f=2+1$ 0.01/0.05 coarse 
ensemble and $aM^0_b=2.8$. Note the large magnification of the 
$y$-axis scale to make the errors visible. 
}\label{fig:csq}\end{figure}

An 
equivalent illustration of the stability 
of the $\Upsilon$ mass is given by the ratio 
of the $p$ and $E$ terms in Equation~\ref{eq:disp}. 
This ratio is called the `speed of 
light' and should have value 1 in our units.
We calculate it as:
\begin{equation}
c^2 = \frac{(\Delta E)^2 + 2M\Delta E}{p^2}.\label{eq:csq}
\end{equation}
where $M$ is determined from above using values 
of $p$ corresponding to $n^2$ = 0 
and 1.
This is a quantity much used as a check of how 
well discretisation errors are controlled in 
actions for light quarks on the lattice~\cite{klassenalford}. 
Even the best light quark actions show discrepancies 
of $c^2$ from 1 of a few percent at typical lattice 
spacing values; some of the poorer actions show 
discrepancies of many percent~\cite{kayaba}. Our results are 
tabulated in Table~\ref{tab:mkin} and plotted in Figure~\ref{fig:csq}.
They show that 
$c^2$ never deviates from 1 by as much as 1\% 
and we can determine it with errors of less than 
1\%. 

Deviations from 1 of $c^2$ would be caused by 
both missing higher order relativistic corrections 
in the NRQCD action and discretisation errors in 
the NRQCD and gluon actions. The terms that give rise 
to these errors have been discussed in Section~\ref{sec:calc}. 
Here we can estimate their effect on $c^2$ by assuming 
that the momentum of the $\Upsilon$ is shared between 
the two quarks equally. The relative error in $\Delta E$ from 
missing radiative corrections to the $(\Delta^{(2)})^2$ 
relativistic correction in Equation~\ref{deltaH}
is then $\alpha_s(p/2)^2/(2M_b^0)^2$. This only 
amounts to 0.5\% even at momentum $(2,2,2)$ on 
the coarse set of ensembles, and 0.4\% on the fine set. 
Similarly missing radiative corrections to the 
discretisation errors, i.e. the effect of setting 
$c_5$ and $c_6$ to 1, amount to no more than 0.5\% 
at momentum $(2,2,2)$ on the coarse set of ensembles,
1\% on the super-coarse set and 0.2\% on the fine set. 
These estimates bear out the results shown in 
Figure~\ref{fig:csq} and emphasise how well 
the NRQCD formalism handles heavy quarks at low non-zero
momenta. This is important not only for extracting 
hadron masses but also for the analysis of form 
factors for $B$ meson semi-leptonic decay~\cite{junko}.


We obtain values for the $\Upsilon$ mass from 
Equation~\ref{eq:mkin} 
using momenta 0 and $n^2=1$. These are given in  
Tables~\ref{tab:mkinsuperc},~\ref{tab:mkinc},~\ref{tab:mkinf} 
for our super-coarse, coarse and fine ensembles 
respectively. 

\begin{table}[t]
\centerline{\begin{tabular}{cc}
\hline
\hline
& $0.0082/0.082$ \\
\hline
$aM_{\Upsilon}$
&8.086(17)
\\
$M_{\Upsilon}$/GeV
&9.25(15)(21)
\\
\hline
\hline
\end{tabular}}
\caption{The $\Upsilon$ mass in lattice units and 
physical units obtained from momenta 0 and $n^2=1$ for 
the super-coarse ensemble. The results in lattice units include 
statistical/fitting errors only; the results in physical units 
include two errors. The first is statistical/fitting, dominated by
that from determining the lattice spacing from the $2S-1S$ splitting. The second error 
is the combined systematic error from the lattice spacing determination. }
\label{tab:mkinsuperc}
\end{table}
\begin{table*}[t]
\centerline{\begin{tabular}{ccccccc}
\hline
\hline
& $0.01/0.05$ & $0.02/0.05$ & $0.03/0.05$ & $0.05/0.05$ & 0.02 &Quenched\\
\hline
$aM_{\Upsilon}$
&5.968(29)
&5.942(29)
&5.935(26)
&5.906(34)
&5.875(27)
&5.855(37)
\\
$M_{\Upsilon}$/GeV
&9.52(16)(11)
&9.54(13)(11)
&10.17(31)(12)
&9.72(28)(11)
&9.82(23)(11)
&9.67(15)(11)
\\
\hline
\hline
\end{tabular}}
\caption{The $\Upsilon$ mass in lattice units and 
physical units obtained from momenta 0 and $n^2=1$ for 
the coarse set of ensembles. The results in lattice units include 
statistical/fitting errors only; the results in physical units 
include two errors. The first is statistical/fitting, dominated by
that from determining the lattice spacing from the $2S-1S$ splitting. The second error 
is the combined systematic error from the lattice spacing determination. }
\label{tab:mkinc}
\end{table*}
\begin{table}[t]
\centerline{\begin{tabular}{cccc}
\hline
\hline
& $0.0062/0.031$ & $0.0124/0.031$ & Quenched\\
\hline
$aM_{\Upsilon}$
&4.218(18)
&4.268(15)
&4.145(12)
\\
$M_{\Upsilon}$/GeV
&9.52(13)(7)
&9.87(12)(7)
&9.63(10)(7)
\\
\hline
\hline
\end{tabular}}
\caption{The $\Upsilon$ mass in lattice units and 
physical units obtained from momenta 0 and $n^2=1$ for 
the fine set of ensembles. The results in lattice units include 
statistical/fitting errors only; the results in physical units 
include two errors. The first is statistical/fitting, dominated by
that from determining the lattice spacing from the $2S-1S$ splitting. The second error 
is the combined systematic error from the lattice spacing determination. }
\label{tab:mkinf}
\end{table}

\subsection{Determining the $b$ quark mass}

{\it The bare $b$ mass}. The values for the $\Upsilon$ masses are in agreement with the 
experimental $\Upsilon$ mass for most of our ensembles, 
indicating that the bare quark masses used for the 
$b$ quarks are good ones. There is particularly close agreement for the 
most important ensembles - the most chiral coarse and fine 
unquenched ensembles. 
Because the calculated $\Upsilon$ mass in physical units is 
proportional to a good approximation to the bare quark 
mass in physical units, shifts of the bare lattice quark 
mass and shifts of the lattice spacing compensate 
each other~\cite{mb} when calculating the bare 
quark mass in physical units. This means that errors in 
the lattice spacing determination or in the 
tuning of the bare quark mass in lattice units do not in fact feed 
through into errors in the bare quark mass in GeV. 
This is demonstrated most clearly if we determine
\begin{equation}
\mbare(a) = a\mbare_{sim}\frac{M_{\Upsilon, expt}}{aM_{\Upsilon,latt}}
\end{equation}
where $a\mbare_{sim}$ is the bare lattice $b$ quark mass 
in lattice units used in the calculation (see Table~\ref{tab:latdata}). 
We tabulate the values of $\mbare(a)$ from this formula for 
each of our lattice spacings in Table~\ref{tab:mbunq}. The errors 
quoted include statistical errors (which range from 0.2\% to 0.5\%) 
as well as systematic errors from relativistic/discretization errors. 
The latter are estimated by noting that 
the binding energy is only 6\% of the $\Upsilon$ mass (as we shall 
see below). The leading errors, from missing radiative corrections
to the leading relativistic/discretization corrections, are expected to be 
20-30\% of that 6\%.
Note that the bare mass, $\mbare(a)$, decreases with decreasing 
lattice spacing, as expected for a running mass. 
The masses from the comparable quenched simulations are only slightly 
higher at 4.52 GeV on the 
coarse ensemble and 4.45 GeV on the fine ensemble. 
We now discuss the conversion of the bare mass first to a perturbative 
pole mass and then to an $\overline{MS}$ mass for the $b$ quark. 

{\it Renormalization of the $b$ quark mass}. There are two 
methods for determining the $b$ pole mass, defined 
perturbatively, from an NRQCD simulation~\cite{mb}. 
Both use perturbative results obtained from 
the heavy quark self-energy~\cite{gulez}. The first uses 
perturbation theory to relate the pole mass 
directly to $\mbare(a)$, defined above: 
\be
M_b^\mathrm{pole} = Z_m\,M_b^0 \equiv \left( 1+C_M\,\alpha_s+\cdots\right)\,M_b^0
\label{mb-pole-eqn}
\ee
where $C_M$ depends on $a\mbare$ and values are given 
in Table~\ref{tab:mbunq}~\cite{gulez}.
The second method uses perturbation theory to relate 
the NRQCD energy of the $\Upsilon$ to the `binding energy' 
of the meson and therefore to the pole mass:
\begin{equation}
M_b^{pole} = \frac{1}{2}(M_{\Upsilon,expt}-(E_{sim} - 2E_0))
\label{eq:eomethod}
\end{equation}
where $E_{sim}$ is the fitted NRQCD energy from $\Upsilon$ correlators 
at zero momentum given in Tables~\ref{tab:SI_E_superc},~\ref{tab:SI_E_coarse},~\ref{tab:SI_E_fine}.
Here $E_0$ is the NRQCD energy of an isolated $b$ quark, 
computed in perturbation theory:
\be
aE_0 = C_{E_0}\alpha_s+\cdots.
\ee
$C_{E_0}$ again depends upon $aM_b^0$ and the appropriate values 
are given in Table~\ref{tab:mbunq}). 

Unfortunately the momentum scales for $\alpha_s$ 
in our two perturbative formulas are quite small. 
Working in the $V$ scheme and using BLM scale fixing~\cite{blm,tadimp} 
an earlier analysis, using a slightly different version 
of NRQCD and an unimproved gluon action, found scales 
of order $0.6/a$ for both the $Z_m$ and $aE_0$ expansions~\cite{colin2}.
Even on our finest lattices, $\alpha_V(0.6/a)$ is about 
0.75~\cite{quentin}. The small scale is expected. 
It reflects real infrared sensitivity in the pole mass. 
Such large values for the coupling constant mean that 
we cannot obtain
an accurate estimate of the pole mass. 

\begin{table*}[t]
\centerline{\begin{tabular}{ccccc}
\hline
\hline
& $M_b^0$/GeV & $C_M$ & $C_{E_0}$ & $M_b^{\overline{MS}}(M_b^{\overline{MS}})$/GeV \\
\hline
super-coarse
&4.68(9)
&0.082
&0.850
&4.31(44)
\\
coarse
&4.44(7)
&0.235
&0.767
&4.30(31)
\\
fine
&4.37(5)
&0.421
&0.689
&4.44(26)
\\
\hline
\hline
\end{tabular}}
\caption{The $b$ quark mass in physical units from 
the most chiral 2+1 flavor lattice ensembles (i.e. 0.0082/0.082 super-coarse, 
0.01/0.05 coarse and 0.0062/0.031 fine). 
The bare mass is given on the left, the perturbative 
coefficients in $Z_m$ and $E_0$ (see text) in the centre 
and the mass in the $\overline{MS}$ scheme at 
its own scale on the right. For the bare mass
the error 
is dominated by a systematic error from the NRQCD action; 
for the $\overline{MS}$ mass the error is an estimate 
from unknown higher orders in perturbation theory. }

\label{tab:mbunq}
\end{table*}

\begin{figure}[t]
{\includegraphics[width=7cm,angle=-90]{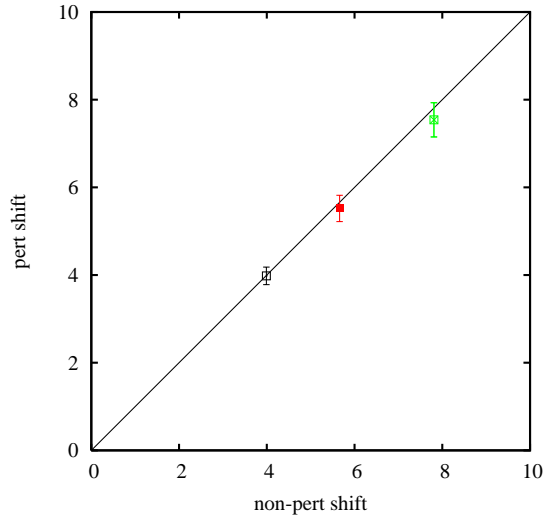}}
\caption{The perturbative estimate of the shift between 
mass and simulation energy for the $\Upsilon$ 
versus the result obtained non-perturbatively from our 
calculations. The results are given in lattice units 
for the most chiral 2+1 flavor ensembles on 
super-coarse (0.0082/0.082), coarse (0.01/0.05) and fine (0.0062/0.031) 
lattices (right to left). 
The errors on the perturbative points are estimates 
of unknown $\alpha_s^2$ terms, taken as $2\times\alpha_V^2(q*)$ for 2 $b$ 
quarks in an $\Upsilon$; the errors on the 
non-perturbative results are negligible in comparison (0.5\%). The 
line drawn is $y=x$ for comparison. }\label{fig:shift}\end{figure}

By equating our two formulas for $M_b^\mathrm{pole}$, 
we obtain an interesting relationship between 
$b$~quark and $\Upsilon$-meson properties:
\be
2( Z_M M_b^0 - E_0) = M_\Upsilon - E_\mathrm{sim}
\ee
The quantities on the right-hand side of this equation 
are measured nonperturbatively in our simulations, 
while the left-hand side is determined from perturbation theory. 
The infrared sensitivity that plagues the pole mass cancels between 
the two terms on the right-hand side, so that scales $q^*$ for $\alpha_V$ vary from 
approximately 1.7/a to 1.3/a from super-coarse to fine 
lattice spacings\,\cite{colinpriv}. 
We compare the perturbative predictions (left-hand side) 
and our nonperturbative measurements (right-hand side) for 
this shift in Figure~\ref{fig:shift}. The agreement is excellent even 
as this shift varies by almost a factor of two.

{\it The $\overline{MS}$ $b$ quark mass}. 
Our inability to obtain a pole mass is not much of a handicap 
since most applications require the $\msb$ mass. 
The $\msb$ mass, like $M_b^0$, is a bare mass, and therefore 
is free of the infrared problems associated with the pole mass. 
We can convert our previous formulas for $M_b^\mathrm{pole}$ into 
formulas for $M_b^{\overline{\mathrm{MS}}}$ using the standard formula,
\be
M_b^{\overline{\mathrm{MS}}}(\mu) = \left(1 - \frac{\alpha_s}{\pi}\,(\frac{4}{3} + 2\log(\frac{\mu}{M_b})+\cdots\right)\,M_b^\mathrm{pole}.
\ee
Combining this formula with Equation~\ref{mb-pole-eqn} gives
\be
M_b^{\overline{\mathrm{MS}}}(\mu) = 
\left( 1 + \frac{\alpha_s}{\pi}\,(C_M\pi -\frac{4}{3} - 2\log(\frac{\mu}{M_b^0}))+\cdots\right)\,M_b^0.
\label{mb-msb-eqn}
\ee
The choice of the $\msb$ scale~$\mu$ is arbitrary, 
but we want to choose a scale that matches the $\msb$ and 
lattice cutoffs\,---\,so that the two bare masses, 
$M_b^{\overline{\mathrm{MS}}}(\mu)$ and $M_b^0$, are roughly 
equivalent. An obvious choice is the scale~$\mu^*$ such that 
the first-order coefficient in Equation~\ref{mb-msb-eqn} vanishes; 
this gives values for $a\mu*$ between 2.4 and 1.9 for our three 
lattice spacings, showing only weak $a$-dependence 
(as expected when $aM_b\gg1$). We simplify the analysis here by 
setting $\mu=2/a$ for all of our lattice spacings, resulting in first-order 
coefficients that range from 0.16 to 0.06. Previous work~\cite{kent2} suggests 
that an appropriate coupling constant in this formula is $\alpha_V(2/a)$, 
where, as expected, the scale is much larger than for the pole-mass formulas. 
We again take values for $\alpha_V$ from~\cite{quentin}.
Our final results, evolved to the $\msb$ mass using 3-loop evolution~\cite{melnikov}, 
are quoted in Table~\ref{tab:mbunq}. These include a perturbative error 
uncertainty which we estimate to be $1\times\alpha_V^2(2/a)$. 
Our results from different lattice spacings are in excellent agreement; 
we take 4.4(3)\,GeV as our final result.

{\it Further checks of the $b$ quark mass}. 
An important check of the $b$ quark mass is whether 
the masses of other hadrons containing $b$ quarks 
calculated on the lattice agree with experiment when 
the $b$ quark mass is fixed from the $\Upsilon$ mass. 
In the quenched approximation this has always been 
problematic as a result of the ambiguity in 
determination of the lattice spacing. 
This led to large differences in $b$ quark masses 
used in $B$ calculations and $\Upsilon$ calculations 
 - 30\% at $\beta$=6.0~\cite{upspaper, arifa}. 
In a lattice simulation which 
matches the real world there is only one value for 
the lattice spacing and one value for the bare $b$ 
quark mass. Here we test this for the MILC 2+1 flavor
configurations. 

\begin{figure}[t]
{\includegraphics[width=7cm,angle=-90]{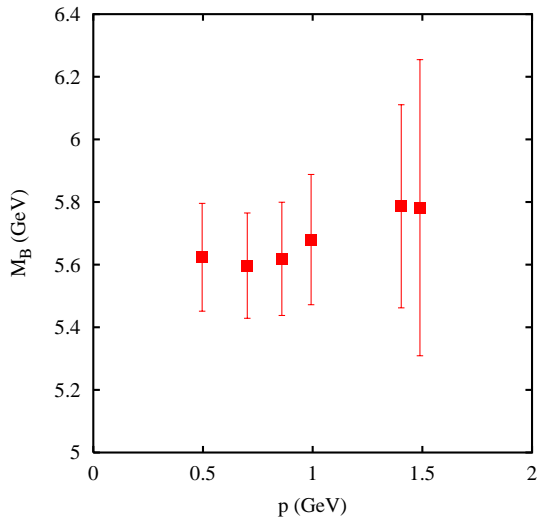}}
\caption{The value of the $B_s$ mass in 
physical units plotted against 
the lattice momentum for the $n_f=2+1$ 0.01/0.05 coarse 
ensemble for $aM^0_b=2.8$, and $am_s = 0.04$ using the 
MILC convention for light quark mass. Errors included are only those 
from determining $M_{kin}$, since the errors from determining 
the scale are the same for all points. }\label{fig:mkinB}\end{figure}

Figure~\ref{fig:mkinB} gives the mass for 
the $B_s$ meson on the 2+1 flavor 0.01/0.05 
coarse ensemble using a bare $b$ quark mass that 
is the same as that of the $\Upsilon$ calculation 
and fixing the lattice spacing from the $2S-1S$ 
$\Upsilon$ splitting, again as in the $\Upsilon$ 
calculation. The $B_s$ mass is obtained 
from the dispersion relation, Equation~\ref{eq:disp}, 
as for the $\Upsilon$. The $s$ quark mass is fixed from 
the $K$ mass~\cite{ourms}. The experimental $B_s$ mass is 5.37 GeV 
so the lattice results are 4\% or 1$\sigma$ high, 
i.e. in agreement within errors.  This is a big 
improvement over the results in the quenched approximation 
and shows that these unquenched configurations 
are at last yielding an unambiguous picture of the 
real world. 

There is a potential systematic error of 
1-2\%, as explained above, coming from the 
NRQCD action in the $\Upsilon$ case. 
As we now explain, this source of error is not 
mirrored in the $B$ system, and so does remain a 
potential systematic difference between the two. 
This can certainly explain the results in 
Figure~\ref{fig:mkinB} if they are not simply 
statistical fluctuation, and would indicate that 
further improvements to the NRQCD action would 
remove any discrepancy. 

The power-counting for the NRQCD action for a heavy-light 
meson is quite different from that of a heavy-heavy meson, 
since it is then powers of 
$1/M_b$ that matter, with $\Lambda_{QCD}/M_b \approx$ 10\% 
for the $B$. All terms through 
second-order ($1/M_b^2$) are included at 
tree-level in the NRQCD action we have used, Equation~\ref{deltaH}, 
and we expect smaller systematic errors than in 
$\Upsilon$ system. Formally the largest systematic 
error comes from radiative corrections to the $\vec{\sigma}\cdot\vec{B}/M_b$ 
term but this term itself does not have a large effect as can be seen 
from 
comparing the $B$ meson hyperfine splitting to the 
$B$ mass. 
The $B$ meson binding energy ($E_{sim}-E_0$) is 
quite a different size, and of different sign, to that 
of the $\Upsilon$. It arises largely from the light 
quark sector and here errors from the improved staggered 
action appear. These are formally $\alpha_s (a\Lambda_{QCD})^2$
which can be estimated to be around 2\%, but applied to the 
binding energy this gives a very small error in the 
$B$ meson mass. 

\begin{figure}[t]
{\includegraphics[width=7cm,angle=-90]{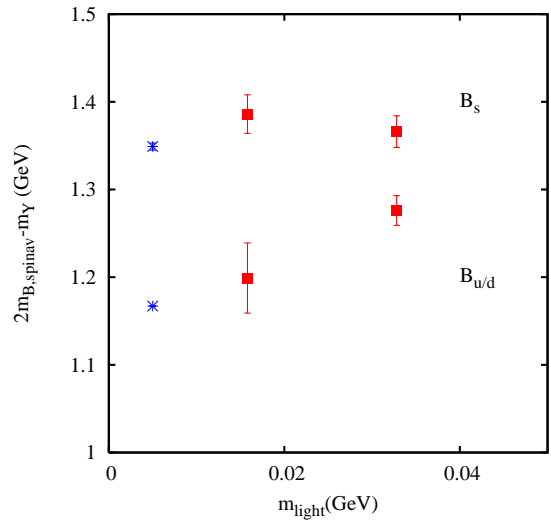}}
\caption[iogth]{The mass difference between twice the spin-average 
of $B$ and $B^*$ and the $\Upsilon$ for $B_{u/d}$ and 
$B_s$, plotted against the sea $u/d$ mass in physical units 
for coarse unquenched configurations. The results 
are for the 2+1 flavor 0.01/0.05 and 0.02/0.05
lattices. The $u/d$ mass used in the $B_{u/d}$ is 
then 0.01 or 0.02 as appropriate and the $s$ mass 
used in the $B_s$ is 0.04, fixed from the $K$ mass~\cite{ourms}. 
The upper points are for the $B_s$ and the lower 
for the $B_{u/d}$. The experimental results are given 
by the bursts~\cite{pdg} where we have assumed that 
the $B^*_s$-$B_s$ splitting is the same as that 
for the $B_{u/d}^*$-$B_{u/d}$. Errors include statistics/fitting 
and $a^{-1}$ errors. 
}\label{fig:Bups}\end{figure}

Another check of agreement with experiment with 
unquenched lattice QCD 
is to study $2m_B - m_{\Upsilon}$ or $2m_{B_s} - m_{\Upsilon}$. 
Explicit terms in the $b$ quark mass cancel in this difference leaving
sensitivity to the differences in binding energies in 
the two systems (with slight heavy quark mass dependence). 
From the discussion above we expect systematic 
errors of order 2\% in this quantity, coming either from 
errors in the NRQCD action on the heavy-heavy side or 
from discretisation errors in the light quark action on 
the heavy-light side. To avoid errors on the heavy-light side 
coming from radiative corrections to the $\vec{\sigma}\cdot\vec{B}$ 
term, we spin-average over pseudoscalar and vector heavy-light 
states. The results are shown in Figure~\ref{fig:Bups} and show 
indeed good agreement with the experimental results at the 
level expected. This is again strong confirmation that 
lattice QCD with light sea quarks reproduces experimental 
hadron masses across a wide range of the spectrum.

\section{Fine structure in the $\Upsilon$ spectrum}\label{sec:fine}
\begin{table}[t]
\centerline{\begin{tabular}{cc}
\hline
\hline
& $0.0082/0.082$ \\
\hline
$aE(1^3S_1)-aE(1^1S_0)$
&0.0348(5)
\\
$aE(2^3S_1)-aE(2^1S_0)$
&0.018(13)
\\
$\frac{E(2^3S_1)-E(2^1S_0)}{E(1^3S_1)-E(1^1S_0)}$
&0.52(38)
\\
\hline
\hline
\end{tabular}}
\caption{$S$-wave fine structure in the $\Upsilon$ spectrum for the super-coarse 
2+1 flavor lattice ensemble. Energies are given in lattice units. 
Errors are statistical/fitting only.}
\label{tab:hyp_E_superc}
\end{table}
\begin{table*}[t]
\centerline{\begin{tabular}{ccccccc}
\hline
\hline
&$0.01/0.05$ & $0.02/0.05$ & $0.03/0.05$ & $0.05/0.05$ & 0.02 &Quenched\\
\hline
$aE(1^3S_1)-aE(1^1S_0)$
&0.03128(34)
&0.03062(32)
&0.02972(33)
&0.02925(43)
&0.02668(34)
&0.02152(30)\\
$aE(2^3S_1)-aE(2^1S_0)$
&0.0111(75)
&0.0172(62)
&0.017(14)
&0.013(13)
&0.0102(109)
&0.0137(78)\\
$\frac{E(2^3S_1)-E(2^1S_0)}{E(1^3S_1)-E(1^1S_0)}$
&0.35(24)
&0.56(20)
&0.57(47)
&0.44(44)
&0.38(41)
&0.64(36) \\
\hline
\hline
\end{tabular}}
\caption{$S$-wave fine structure in the $\Upsilon$ spectrum for the coarse ensembles. 
Energies are given in lattice units. Errors are statistical/fitting only.}
\label{tab:hyp_E_coarse}
\end{table*}
\begin{table}[t]
\centerline{\begin{tabular}{cccc}
\hline
\hline
& $0.0062/0.031$ & $0.0124/0.031$ & Quenched\\
\hline
$aE(1^3S_1)-aE(1^1S_0)$
&0.02599(53)
&0.02422(45)
&0.01735(29)\\
$aE(2^3S_1)-aE(2^1S_0)$
&0.0141(55)
&0.0120(42)
&0.0107(31)\\
$\frac{E(2^3S_1)-E(2^1S_0)}{E(1^3S_1)-E(1^1S_0)}$
&0.54(21)
&0.50(18)
&0.62(18) \\
\hline
\hline
\end{tabular}}
\caption{$S$-wave fine structure in the $\Upsilon$ spectrum for the fine lattice ensembles. Energies are given in lattice units. Errors are statistical/fitting only.}
\label{tab:hyp_E_fine}
\end{table}

\begin{figure}[t]
\includegraphics[width=7cm,angle=-90]{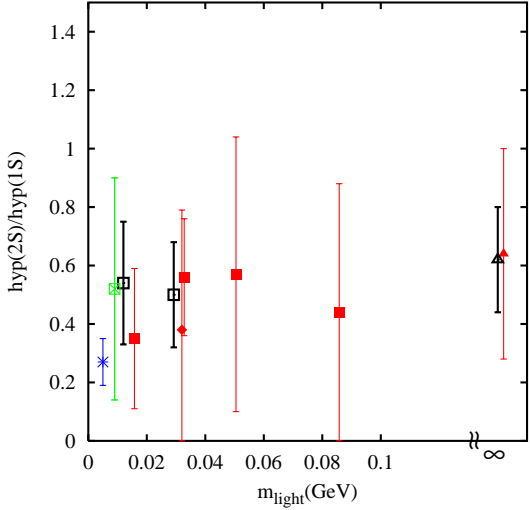}
\caption{The ratio of the splitting between the $\Upsilon^{\prime}$ and 
$\eta^{\prime}_b$ to that between the $\Upsilon$ and the $\eta_b$. 
Crossed squares are from super-coarse lattices, 
closed squares are from coarse lattices and open squares from fine lattices. 
The closed diamond is from the coarse $n_f=2$ ensemble. The closed and open 
triangles are from the coarse and fine quenched ensembles.  
The burst represents the current experimental result for the $\psi$ system, 
using the value for the mass of the $\eta_c^{\prime}$ from the 
BELLE collaboration~\cite{belle}. }
\label{fig:hyprat}
\end{figure}

\begin{figure}[t]
\includegraphics[width=7cm,angle=-90]{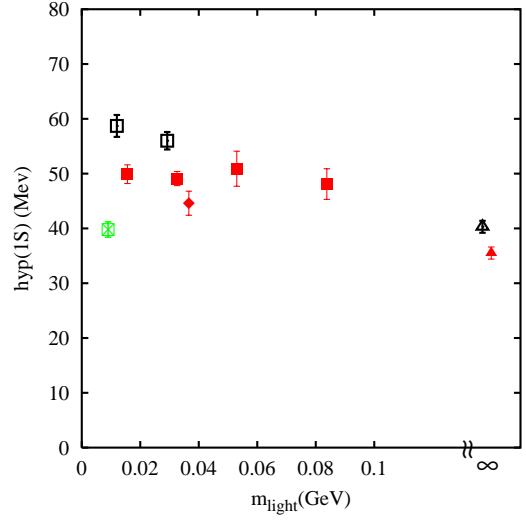}
\caption{The ground-state hyperfine splitting in the $\Upsilon$ system 
as a function of the light quark mass included in the quark vacuum 
polarization. 
Crossed, closed and open symbols are from super-coarse, 
coarse and fine lattices respectively. 
Squares and triangles denote unquenched and quenched results respectively. 
The closed diamond is from the coarse $n_f = 2$ ensemble. 
Errors include both statistical/fitting errors from 
determining the splitting and 
from determining the lattice spacing scale. This latter 
error appears doubled (see text). Note that there is an 
additional overall systematic error of 25\% discussed in the text.}
\label{fig:hypmq}
\end{figure}

\begin{figure}[t]
\includegraphics[width=7cm,angle=-90]{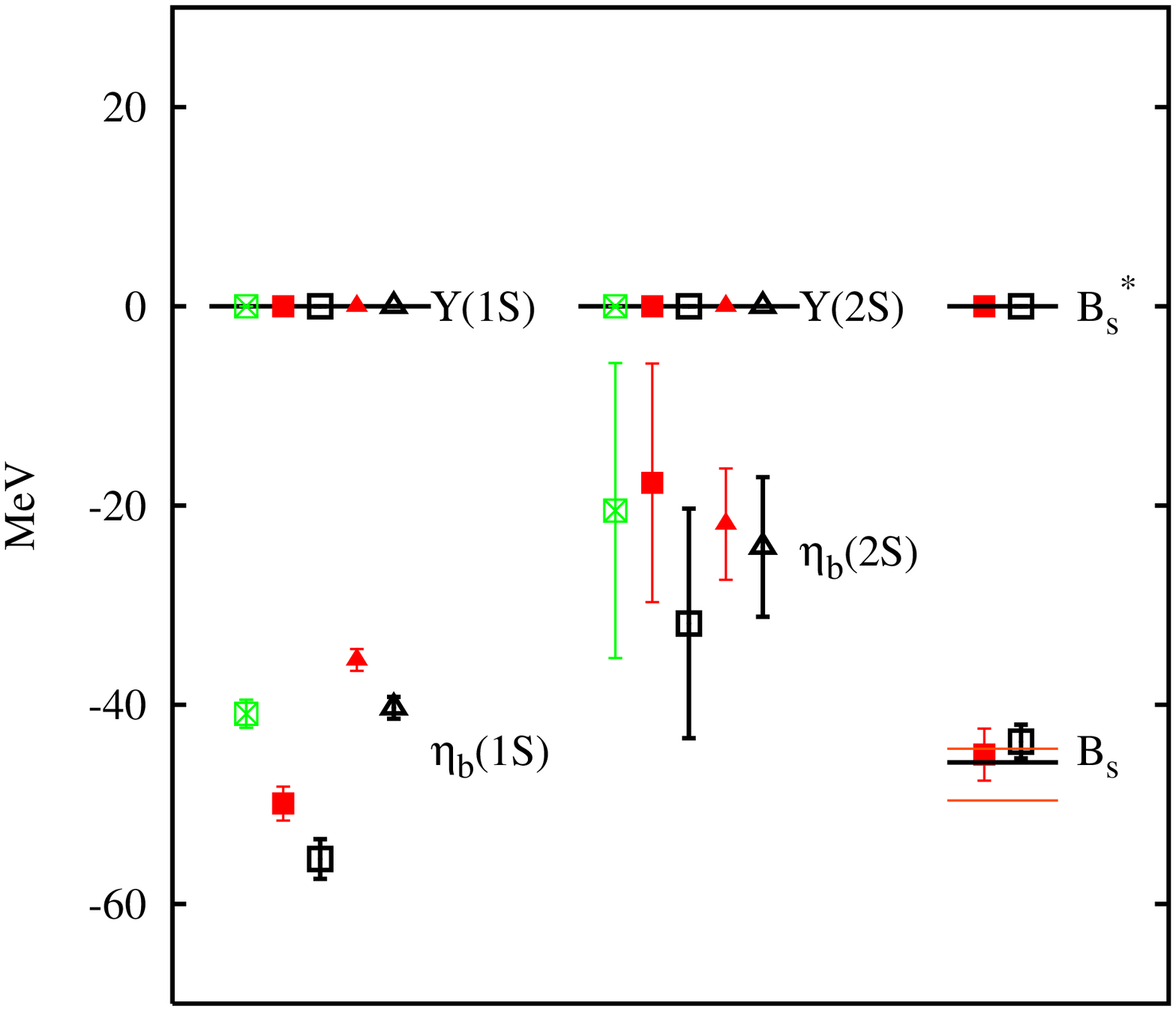}
\caption{Hyperfine splittings in the $\Upsilon$ and $B_s$ systems. 
Crossed, closed and open symbols are from super-coarse, 
coarse and fine lattices respectively. 
Squares and triangles denote unquenched and quenched results respectively. 
The unquenched results are from the ensembles in Table~\ref{tab:latdata} 
that include the lightest light quark mass in the quark vacuum 
polarization at each value of the lattice spacing. 
Errors include both statistical/fitting errors from 
determining the splitting and 
from determining the lattice spacing scale as in Figure~\ref{fig:hypmq}. 
Additional systematic errors are 25\% and discussed 
in the text. 
The unquenched results include a renormalisation to 
the correct value for $u_{0L}$ (see text). 
The faint lines give the experimental limits on the 
$B_s^*-B_s$ splitting and the dark line gives the experimental 
result for $B^*-B$~\cite{pdg}.}
\label{fig:hypspec}
\end{figure}

Fine structure in the $\Upsilon$ spectrum arises from terms in the action 
which are suppressed by a power of $v^2$ compared to the leading term, $H_0$. 
Such sub-leading terms themselves have sub-leading corrections which we 
have not included. We expect therefore that the results we obtain are 
only accurate to $\cal{O}$$(v^2)=10\%$. In addition there are 
$\cal{O}$$(\alpha_s)$ corrections to the coefficients of these terms, 
such as $c_4$ in equation~\ref{deltaH}, which could amount to 
a 20-30\% error. In fact, as we shall see, we are able to do rather 
better than this, but results in this section are still rather qualitative. 
We present them nevertheless as providing useful pointers for 
future calculations. 

\subsection{$S$-wave fine structure}
   
There are no experimental results for either the ground or 
radially excited $S$-wave spin splitting (hyperfine splitting) 
in the $\Upsilon$ system, and so these are key results to be predicted by lattice 
QCD. This splitting is the mass difference between the vector 
($\Upsilon$) state and the pseudoscalar ($\eta_b$) and so 
a prediction of the splitting is equivalent to a prediction 
of the unknown $\eta_b^{(\prime)}$ mass. Our result for the 
ground-state hyperfine splitting is 61(14) MeV, 
giving $M_{\eta_b}$ = 9.399(14) GeV. Our prediction for 
the radially excited hyperfine splitting is 30(19) MeV, giving 
$M_{\eta_b^{\prime}}$ = 9.993(19) GeV. The largest contribution 
to the error for the ground-state splitting 
comes from unknown radiative corrections to the 
spin-dependent term in the NRQCD action that gives rise to this 
splitting. For the excited state splitting the dominant error 
is statistical. How we estimate the size of the errors and how 
we can reduce them in future 
is described in more detail below.  

Results for the ground and radially excited hyperfine 
splitting in lattice units, and 
their ratio are given in Tables~\ref{tab:hyp_E_superc},~\ref{tab:hyp_E_coarse} 
and~\ref{tab:hyp_E_fine} for the super-coarse, coarse and fine MILC lattices respectively. 
Splittings are obtained from the difference in fitted energies of 
the $^3S_1$ ($\Upsilon$) and $^1S_0$ ($\eta_b$) states 
using $3 \times 3$ fits ($2 \times 2$ fits on the super-coarse 
lattices). Bootstrapping the fits together 
improved the errors in some cases, and not in others, and so we
do not give bootstrap errors here. The results for the hyperfine 
splitting are very precise because it is the difference between 
two ground $S$-wave states which are very well-determined. 
This means that statistical errors are small and 
differences between different lattice 
calculations show up clearly. The possibility of 
significant systematic errors in the splitting, however, must not be forgotten.

Figure~\ref{fig:hyprat} shows the ratio of the hyperfine splitting 
for the 1st radial excitation to that of the ground state i.e. 
$(M_{\Upsilon^{\prime}}-M_{\eta_b^{\prime}})/(M_{\Upsilon}-M_{\eta_b})$.
The hyperfine splitting is controlled by the ${\bf \sigma}\cdot{\bf B}$ 
term in the NRQCD action and is then proportional to the square of $c_4$ 
(see Equation~\ref{deltaH}). 
$c_4$ cancels in the ratio so in principle it can be accurately 
determined in our lattice calculation with $c_4$ set to 1. 
The statistical errors here are large but we obtain a result of 
0.5(3). 
There is no sign of significant dependence on 
the lattice spacing or on the light quark mass.  
The result is in 
keeping with the experimental results in the charmonium system, plotted 
as a burst in the Figure~\cite{belle}. The charmonium result 
is not irrelevant because 
we also expect a lot of the heavy quark mass dependence to cancel 
in the ratio of splittings. 

Figure~\ref{fig:hypmq} shows our results for the ground-state 
hyperfine splitting in MeV as a function of the bare quark 
mass of the light quarks included in the quark vacuum polarisation. 
We use the lattice spacing values from the $2S-1S$ splitting 
of Section~\ref{sec:scale} to convert to physical units. 
Results are shown for the super-coarse, coarse and fine ensembles.
The errors on the points include the very small statistical 
error from the splitting itself and also the much larger 
error from the uncertainty in the lattice spacing. This uncertainty
in fact appears multiplied by two because the hyperfine splitting 
is approximately inversely proportional to the physical quark 
mass. Any underestimate of the inverse lattice spacing will 
both reduce the hyperfine splitting directly and also 
reduce it indirectly by reducing the quark mass~\cite{scaling}, 
so that the splitting must then be readjusted downwards to correspond 
to the correct meson mass. 
We then have a statistical/fitting error of 5\% on the ground-state
hyperfine splitting 
in physical units. 

The results show significant dependence on both the lattice 
spacing (to be discussed further below) and the number of 
flavors of quark, $n_f$, included in the quark vacuum polarisation at 
a given lattice spacing. Neither of these points is surprising.
In the language of a simple potential model the hyperfine splitting arises 
from a very short-range spin-spin potential, dominated by 
the perturbative contribution which 
is a delta function at the origin. 
We therefore expect sensitivity to the lattice spacing~\cite{scaling} and 
also to the value of $\alpha_s$ at short distance, which 
depends strongly on $n_f$~\cite{quentin}. The results do not 
however, show significant dependence on the bare light 
quark mass. This is also not surprising, as for spin-independent 
splittings, because of the 
large gluon momenta involved. 
Provided that the light quark mass is light enough, as is 
the case here, the splitting 
simply `counts' the number of light quark flavors. 
We do not therefore attempt any extrapolation in the light 
quark mass. 

Further examination of the dependence on the lattice spacing 
of the $2+1$-flavor results is necessary, however, to obtain 
a result relevant to the real world of continuous space-time. 
It is important to disentangle discretisation errors 
from errors resulting from unknown radiative corrections in 
$c_4$. One important issue here is tadpole-improvement.
We expect that radiative corrections to $c_4$ are not large if we use 
tadpole-improved gauge fields. We can use any sensible 
quantity to determine the tadpole-improvement factor, $u_0$. 
The results will be independent of the quantity used in a 
complete calculation because 
any change will be compensated by corrections to $c_4$. Here, 
however, we have not included any corrections to $c_4$. 
We have also made slight errors in our determination of 
$u_0$ from the average link in Landau gauge on the fine 
and super-coarse lattices. Because of the large number of 
gauge fields appearing in the lattice discretisation of 
the $B$ field in the $\sigma \cdot B$ term, the hyperfine 
splitting is very sensitive to $u_0$. It is proportional 
to $u_0^{-6}$, when the renormalisation of the quark 
mass is taken into account~\cite{scaling}. Subsequent 
more accurate determination of $u_{0L}$ has shown that 
it has the value 0.8541 on the 2+1-flavor 0.0062/0.031 fine 
ensemble (i.e. 1\% higher than in Table~\ref{tab:latdata}) 
and 0.8102 on the 2+1-flavor 0.0082/0.082 super-coarse 
ensemble (i.e. 0.5\% lower than in Table~\ref{tab:latdata}). 
We then adjust our hyperfine splitting results on these 
two ensembles by multiplying by the ratio of the 
sixth power of the $u_{0L}$ value used in the simulation 
to the correct value above, before studying the results 
as a function of lattice spacing. 

Figure~\ref{fig:hypspec} shows the results for the 
hyperfine splitting on the unquenched 
ensembles with lightest $u/d$ quark mass with this 
renormalisation in place. It also shows again the 
quenched results for comparison. The discretisation 
effects seen are consistent with errors proportional to $a^2$ 
in the form: 
\begin{equation}
{\rm Hyperfine}(a) = {\rm Hyperfine}(0)(1-(\Lambda a)^2 + \ldots).
\label{eq:hypvsa}
\end{equation}
We find $\Lambda \approx$ 640 MeV. This variation with lattice spacing is 
sizeable, but smaller than previous results using an 
unimproved gluon action and an unimproved operator for the $B$
field~\cite{scaling}. Using a constrained fitting 
method with multiple higher order terms~\cite{bayes} 
we obtain a value for the hyperfine splitting in 
the continuum including 2+1 flavors of light quarks of 
61(4) MeV where the 7\% error now includes both the 
statistical/fitting errors on the individual points and 
the uncertainty in the continuum value from discretisation 
errors. It is dominated by the latter error.

We now estimate the size of the error from unknown radiative 
corrections to $c_4$, appropriate to tadpole-improvement 
with $u_{0L}$. In principle this should be done before 
studying the discretisation errors above but, as we shall 
see, the $c_4$ effects are not dependent on the lattice spacing 
within the errors that we have so we have decoupled the two effects for clarity. 
Radiative corrections to $c_4$ are in principle $\cal{O}$$(\alpha_s)$ 
and this would give an error of approximately 20-30\% here. 
However, we can make use of non-perturbative information 
from the experimental spectrum of heavy-light mesons 
to limit this uncertainty.

Figure~\ref{fig:hypspec} also shows our results for the 
hyperfine splitting between the vector ($B_s^*$) and 
pseudoscalar ($B_s$) ground states of the $b\overline{s}$ system, 
compared to experiment~\cite{newB}. The experimental results for this 
splitting have significant errors, but results from the 
$c$-light mesons show very little difference between the 
hyperfine splitting for $D_s$ and $D$. We therefore mark 
also on this figure the result for the $B^*-B$ splitting 
(45.8(4) MeV)
in the expectation that the $B_s^*-B_s$ splitting will 
in fact be very close to this. 
Our lattice results are obtained by studying correlators 
of NRQCD $b$ quarks and improved staggered $s$ quarks. The 
$b$ quarks have exactly the same NRQCD action and bare quark 
mass as that described here for the $\Upsilon$ system. We
used the correct value for $u_{0L}$ on the fine unquenched 
lattices. The 
bare $s$ quark used for the valence $s$ quarks 
(in lattice units and using the MILC 
convention) is 0.04 on the coarse lattices and 0.031 on the 
fine lattices. The value on the coarse lattices is lower 
than that included in the quark vacuum polarization (see Table~\ref{tab:latdata}), 
but more closely matches the correct value of the $s$ quark 
mass determined after the simulation~\cite{ourms}.

The lattice result for the $B_{(s)}^*-B_{(s)}$ splitting  
is also sensitive to $c_4$, but only linearly since 
the system has only one $b$ quark. We can then use the 
results of Figure~\ref{fig:hypspec} to determine $c_4$ 
non-perturbatively. Using just the lattice $B_s$ results 
we would conclude that $c_4$=1 on both coarse and 
fine lattices up to 5\% errors from 
both the lattice calculation and the experimental result. 
We can also study the $B^*-B$ splitting on the lattice, 
but here our results are somewhat less precise~\cite{newB}. The 
central value on some ensembles is as much as 10\% below 
experiment, which would indicate the need for $c_4$=1.1. 
On this basis we take a non-perturbative result for 
$c_4$ of 1.0 but allow for a 10\% error.  
This gives then an additional 20\% error from $c_4^2$ 
on the $\Upsilon$ hyperfine splitting. 

Our final result for the ground state hyperfine splitting 
is 61(4)(12)(6) MeV where the errors are: statistical/fitting and
discretisation errors; radiative corrections and relativistic 
corrections (which we take to be 10\%) respectively.
Using the ratio, this gives a prediction for the 
radially excited splitting of 30(19) MeV where the 
dominant error is statistical. 
The central value of the ground-state splitting is larger than 
previous calculations using $n_f$ = 2 flavors of sea
quarks~\cite{oldalpha, sesam, marcantonio, manke} which 
range from 20 to 50 MeV.
As well as not having the correct value of $n_f$, these older 
results do not have access to the range of lattice spacing 
values that we have here or to the light quark masses in the quark 
vacuum polarisation.
Some of them~\cite{oldalpha, marcantonio} use unimproved 
gluon actions and an unimproved $B$ field in the NRQCD 
action. These results then have even larger discretisation 
errors than here that tend to suppress the hyperfine 
splitting further~\cite{scaling}. 
Some of the calculations~\cite{sesam, manke} do include 
improved $B$ fields as here, and higher order spin-dependent terms 
in the NRQCD action as well. Their results 
indicate that the inclusion of the higher-order terms tend to reduce the hyperfine 
splitting~\cite{trottier}. However, this is without any determination 
of $c_4$ which is at least as large an effect. 
The new result that we give here is then more realistic than 
previous calculations and with a more reliable error.  Future 
calculations need to include radiative corrections to 
$c_4$ and relativistic corrections to spin-dependent terms. 
A more extensive analysis of discretisation effects will 
then be possible. 

\begin{table}[t]
\centerline{\begin{tabular}{cc}
\hline
\hline
&$0.0082/0.082$ \\
\hline
$1^3P_1-1^3P_0$
&0.018(14)
\\
$1^3P_{2E}-1^3P_1$
&0.016(15)
\\
$1^3P_{2E}-1^3P_{2T}$
&0.005(16)
\\
$\frac{1^3P_1-1^3P_0}{1^3P_2-1^3P_1}$
&1.1(1.3)
\\
$1^3\bar{P}-1^1P_1$
&0.02(1)
\\
\hline
\hline
\end{tabular}}
\caption{$P$-wave fine structure for the super-coarse 2+1 flavor ensemble. Energies are 
given in lattice units.  
Errors are statistical/fitting only and obtained by directly from the fits.}
\label{tab:pfine_superc}
\end{table}
\begin{table}[t]
\centerline{\begin{tabular}{cc}
\hline
\hline
&$0.01/0.05$ \\
\hline
$1^3P_1-1^3P_0$
&0.0168(20)
\\
$1^3P_2-1^3P_1$
&0.0139(21)
\\
$\frac{1^3P_1-1^3P_0}{1^3P_2-1^3P_1}$
&1.20(27)
\\
$1^3\bar{P}-1^1P_1$
&0.01(1)
\\
\hline
\hline
\end{tabular}}
\caption{$P$-wave fine structure for the 2+1 flavor 0.01/0.05 coarse ensemble. Energies are 
given in lattice units. Note that these results use the half the 
number of configurations of other results because of a loss of data. 
Errors are statistical/fitting only and obtained by bootstrap. }
\label{tab:pfine_coarse}
\end{table}
\begin{table}[t]
\centerline{\begin{tabular}{ccc}
\hline
\hline
& $0.0062/0.031$ & Quenched\\
\hline
$1^3P_1-1^3P_0$
&0.0142(34)
&0.0096(20)\\
$1^3P_2-1^3P_1$
&0.0046(42)
&0.0076(20)\\
$\frac{1^3P_1-1^3P_0}{1^3P_2-1^3P_1}$
&3.1(2.8)
&1.27(57)\\
$1^3\bar{P}-1^1P_1$
&0.0006(20)
&-0.0002(14)\\
\hline
\hline
\end{tabular}}
\caption{$P$-wave fine structure for two of the fine lattice ensembles, quenched and 
unquenched. Energies are 
given in lattice units. Errors are statistical/fitting only and obtained 
by bootstrap. }
\label{tab:pfine_fine}
\end{table}

\begin{figure}[t]
\includegraphics[width=7cm,angle=-90]{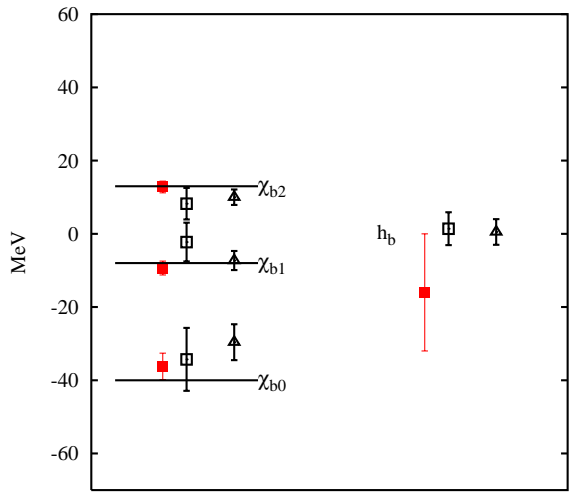}
\caption{$P$-wave spin splittings. Closed and open symbols 
are from coarse and fine lattices respectively. 
Squares and triangles denote unquenched and quenched 
results respectively. 
The unquenched results are from the ensembles in Table~\ref{tab:latdata} 
that include the lightest light quark mass in the quark vacuum 
polarization at the coarse and fine lattice spacing values. 
Errors are statistical/fitting only but 
include errors from determining the lattice spacing as in Figure~\ref{fig:hypmq}. 
There is an additional 25\% systematic error as discussed 
in the text. Lines give the experimental results where available~\cite{pdg}.}
\label{fig:sdspec}
\end{figure}

\subsection{$P$-wave fine structure}

Both ground and radially excited $^3P_{0,1,2}$ states ($\chi_{bJ}$ and $\chi_{bJ}^{\prime}$
are well measured in the $\Upsilon$ system. No $^1P_1$ ($h_b^{(\prime)}$)
states have been seen. We are able to predict that the ground 
state $h_b$ should lie within 6 MeV of the spin-average of the 
$\chi_b$ states i.e. at 9900(3)(6) MeV, where 3 is the experimental 
error~\cite{pdg}. We describe our analysis in more detail below, 
including a discussion of how well our $^3P$ results match experiment. 

The $P$-wave fine structure was only calculated for a limited 
selection of ensembles. For the majority of runs, we were unable to 
resolve excited splittings well and consequently show only the ground 
state results. 
The errors given are those from bootstrapping the fits of 
the individual correlators on the coarse and fine ensembles. For 
the super-coarse ensembles they come directly from the fits. 
Tables~\ref{tab:pfine_superc},~\ref{tab:pfine_coarse},~\ref{tab:pfine_fine} 
give the energy splittings in lattice units.
The lattice representation for the $^3P_2$ 
states used was the $E$ representation on the coarse and fine 
lattices. Any difference between the energy in the 
$E$ representation and in the $T$ representation is a 
lattice artefact. We give both for the super-coarse 
ensemble where such artefacts will be largest, but 
even there we cannot resolve any difference. 
The Tables also give
the
dimensionless ratio: 
\beq
\frac{M(1^3P_1)-M(1^3P_0)}{M(1^3P_2)-M(1^3P_1)}.
\label{}
\eeq 
The results obtained for this ratio have sizeable errors but 
are consistent on both unquenched and quenched ensembles with the 
experimental value of 1.65(9)~\cite{pdg}. 

Figure~\ref{fig:sdspec} plots the lattice results for the ground 
state $P$-wave splittings, each relative to the spin average. 
We use the $2S-1S$ lattice spacing values to convert to 
physical units. 
The plot shows the coarse and fine unquenched and fine quenched 
results only since the results on the super-coarse ensemble 
have error bars which are too large to be useful, although 
they are consistent with the others. The Figure shows that 
the size of the quenched $P$-wave splittings are 
smaller than experiment. The results on the coarse 2+1 flavor
configurations look consistent with experiment, as do the 
fine 2+1 flavor results, although their errors are rather large. 
The errors shown on the plot are from statistics/fitting only 
and from determining the lattice spacing. 
We double the lattice spacing error as for the 
hyperfine splitting but in general this is negligible compared 
to the statistical/fitting error of the $P$-wave splittings 
themselves. We now discuss the sources of 
systematic error. 

As for the hyperfine splitting, the largest sources of 
systematic error potentially are the unknown radiative 
corrections to relevant spin-dependent terms in the 
NRQCD action. Two terms are relevant for $P$-wave fine 
structure which complicates the picture - the $\sigma\cdot B$ 
term and the $\sigma\cdot (D\times E)$ term - with 
coefficients $c_4$ and $c_3$ respectively. There are combinations 
of $P$-wave splittings, however, that can be compared to experiment 
to give non-perturbative information to fix these 
coefficients. 

The $\sigma\cdot (D\times E)$ term gives rise to a coupling 
between spin and orbital angular momentum and a term in 
the mass of each state which depends on the expectation 
value of $L \cdot S$ multiplied by $c_3$. 
The $\sigma\cdot B$ term gives rise 
to a mass term proportional to $c_4^2$ and either $S \cdot S$, 
which is the same for all $^3P$ states, or the combination 
$3(S\cdot\hat{n}S\cdot\hat{n})-S^2$ which does distinguish 
between $^3P$ states. From the expectation values of these 
operators in the different $\chi_b$ states~\cite{cdschladming} we see that the 
combination 
\begin{equation}
5M(\chi_{b2}) - 3M(\chi_{b1}) - 2M(\chi_{b0})
\end{equation}
is proportional to $c_3$ and independent of $c_4$. 
The experimental result for this combination is 
165(4) MeV and our lattice result on the coarse 
unquenched ensemble is 164(16) MeV. This indicates that 
$c_3$ is 1.0 to within an error of 10\%. 
The combination 
\begin{equation}
M(\chi_{b2}) - 3M(\chi_{b1}) + 2M(\chi_{b0})
\end{equation}
is proportional to $c_4^2$ and independent of $c_3$. 
The experimental result for this combination is 
-46(3) MeV and our lattice result is -31(7) MeV. The 
lattice result is 30\% low but with a large error, so 
that the difference is 2$\sigma$. 
The limits on $c_4^2$ from the $B_s$ hyperfine splitting, 
i.e. 1.0 with an error of 20\%, are to be preferred for 
accuracy. 

We conclude from this that the systematic error from 
radiative corrections to the leading spin-dependent terms 
is $\cal{O}$(20\%) in the $^3P$ fine structure. 
We also expect a 10\% error from higher order relativistic 
corrections. Discretisation errors are not expected 
to be as severe here as for the hyperfine splitting because 
the $P$-wave states have a larger radial extent than the 
$S$-wave and, in potential model language, the potentials 
that give rise to the relevant splittings are not so 
concentrated at the origin. Allowing for as much as a 10\% error 
from discretisation effects, however, the total expected systematic 
error in the $^3P$ fine structure in this calculation is 25\%. 
Future calculations will include radiative and relativistic 
corrections and will then resolve the $^3P$ fine structure 
much more accurately. 

Figure~\ref{fig:sdspec} also shows the splitting between 
the spin-average of the $^3P$ states and the $^1P_1$. 
This splitting is a result of the same $S\cdot S$ coupling 
that produces the hyperfine splitting but it is expected to
vanish for $P$-wave states in a potential model because 
the interaction is focussed at the origin. We have 
no signal for such a splitting with an error of 5 MeV on 
the fine unquenched ensemble from statistics/fitting and 
the determination of the lattice spacing. Systematic 
errors arise from radiative corrections to $c_4^2$, which 
we take to be 20\% as above, and relativistic and discretisation 
errors which we take to be 10\% each. This gives a 
splitting of zero with an error of 6 MeV. 

\begin{table}[t]
\centerline{\begin{tabular}{cc}
\hline
\hline
&$0.01/0.05$ \\
\hline
$1^3D_3-1^3D_2$
&0.010(15)
\\
$1^3D_2-1^1D_2$
&0.002(18)
\\
\hline
\hline
\end{tabular}}
\caption{$D$-wave fine structure for the 2+1 flavor 0.01/0.05 coarse ensemble. 
Energies are 
given in lattice units. 
Errors are statistical/fitting only and obtained directly from the fits.}
\label{tab:dstates}
\end{table}

\subsection{$D$-wave fine structure}

The fine structure in the $D$-wave sector is on a smaller 
scale than for $S$-wave and $P$-wave, as expected. We are unable to resolve any 
splittings there with errors of around 10 MeV. $D$-wave 
fine structure was only studied on the 2+1 flavor 0.01/0.05 
coarse ensemble. We looked at the $^1D_2$ state 
(using the $T$ representation), the $^3D_2$ (using the 
$E$ representation) and the $^3D_3$ (using the 
$A$ representation)~\cite{upspaper}. Table~\ref{tab:dstates} 
gives splittings in lattice units with errors 
obtained directly from the fits. 

\section{The $\Upsilon$ leptonic width}\label{sec:lepwid}

The width of $\Upsilon$ decay to two leptons, $\Gamma_{ee}$, 
can be well measured
experimentally~\cite{pdg, cleolw}. The matrix element that 
gives rise to the decay can also be calculated 
on the lattice and we describe that calculation here. 
Our most accurate result is for the ratio between 
the $\Upsilon^{\prime}$ and $\Upsilon$ of $\Gamma_{ee}M_{\Upsilon (nS)}^2$. 
We obtain 0.48(5) compared to the recent preliminary experimental 
result of 0.517(10)~\cite{cleolw}. Below we describe the 
calculation in more detail and the sources of error, along 
with improvements that will shortly be possible in the result. 

On the lattice $\Gamma_{ee}$ is calculated from 
the matrix element of the $b\overline{b}$ vector current  
between the $\Upsilon$ and the vacuum.  The decay width is 
then given by~\cite{perkins}:
\beq                                                                           
 \Gamma_{ee}(nS)=16\pi\alpha_{em}^2e_b^2\frac{\langle \Upsilon_n | J_v | 0 \rangle^2}{6M^2_{\Upsilon(nS)}}({Z_{match}})^2
\label{eq:gammaeepsi}                                                          
\eeq  
where $\alpha_{em}$ is the electromagnetic coupling constant, $e_b$ 
is the charge on a $b$ quark in units of the charge on an electron and 
$M_{\Upsilon(nS)}$ is the mass of the $n$th radial excitation of 
the $\Upsilon$. This formula assumes a non-relativistic normalisation 
of states, $\langle \Upsilon | \Upsilon \rangle$ = 1.  
$Z_{match}$ is the renormalisation constant, $1+z^{(1)}\alpha_s + \ldots$, 
required to match the lattice vector current to a continuum renormalisation 
scheme. The lattice vector current constructed out of 
NRQCD fields has leading term $\chi^{\dag}_b{\bf{\sigma}}\psi_b$ where 
$\psi$ annihilates a $b$ quark and $\chi^{\dag}$ an anti-$b$. 
There are sub-leading terms which take account of relativistic 
and discretisation corrections required to match 
the continuum current to higher order. This matching can be 
done perturbatively and is in progress~\cite{horgan}. The 
calculation is entirely analogous to that for the $B$ meson 
decay constant, whose $Z_{match}$ has been calculated 
through 1-loop~\cite{junkocolin}, including 
the effect of higher order current corrections. 

\begin{table}[t]
\centerline{\begin{tabular}{cc}
\hline
\hline
& $0.0082/0.082$ \\
\hline
$a^{3/2}\Psi_1(0)$
&0.5184(10) \\
$a^{3/2}\Psi_2(0)$
&0.499(20)\\
\hline
\hline
\end{tabular}}
\caption{Results for the leading order piece of the lattice 
heavy-heavy vector current matrix element between the $\Upsilon$
and the vacuum, also known as the `wavefunction at the origin'.
Results are given in lattice units for the $\Upsilon$ ($\Psi_1(0)$)
and for the $\Upsilon^{\prime}$ ($\Psi_2(0)$) for the 
super-coarse ensembles. Errors are statistical/fitting only. } 
\label{tab:wfnsuperc}
\end{table}
\begin{table*}[t]
\centerline{\begin{tabular}{ccccccc}
\hline
\hline
& $0.01/0.05$ & $0.02/0.05$ & $0.03/0.05$ & $0.05/0.05$ & 0.02 &Quenched\\
\hline
$a^{3/2}\Psi_1(0)$
&0.30149(66)
&0.29830(59)
&0.29441(67)
&0.29142(78)
&0.28563(53)
&0.26152(50)\\
$a^{3/2}\Psi_2(0)$
&0.2446(66)
&0.2401(50)
&0.224(14)
&0.237(11)
&0.230(11)
&0.2346(48)\\
\hline
\hline
\end{tabular}}
\caption{Results for the leading order piece of the lattice 
heavy-heavy vector current matrix element between the $\Upsilon$
and the vacuum, also known as the `wavefunction at the origin'.
Results are given in lattice units for the $\Upsilon$ ($\Psi_1(0)$)
and for the $\Upsilon^{\prime}$ ($\Psi_2(0)$) for the 
coarse ensembles. Errors are statistical/fitting only. } 
\label{tab:wfncoarse}
\end{table*}
\begin{table}[t]
\centerline{\begin{tabular}{cccc}
\hline
\hline
& $0.0062/0.031$ & $0.0124/0.031$ &Quenched\\
\hline
$a^{3/2}\Psi_1(0)$
&0.17666(49) 
&0.17796(44) 
&0.15159(31) \\
$a^{3/2}\Psi_2(0)$
&0.1357(24)
&0.1372(19)
&0.1298(17)\\
\hline
\hline
\end{tabular}}
\caption{Results for the leading order piece of the lattice 
heavy-heavy vector current matrix element between the $\Upsilon$
and the vacuum, also known as the `wavefunction at the origin'.
Results are given in lattice units for the $\Upsilon$ ($\Psi_1(0)$)
and for the $\Upsilon^{\prime}$ ($\Psi_2(0)$) for the 
fine ensembles. Errors are statistical/fitting only. } 
\label{tab:wfnfine}
\end{table}

\begin{figure}[h]
\includegraphics[width=7cm,angle=-90]{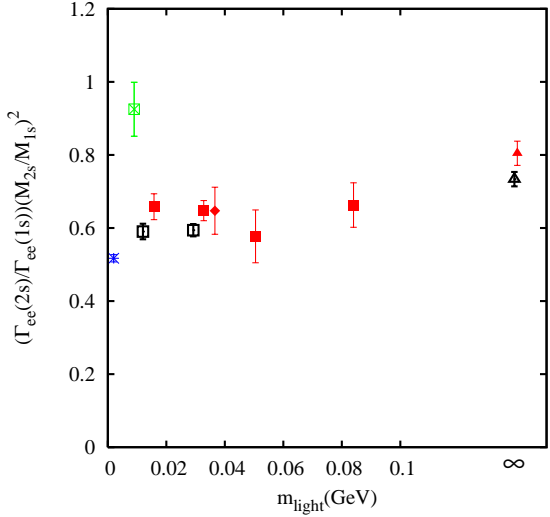}
\caption{The ratio of leptonic width times mass squared for the 1st excited state 
of the $\Upsilon$ to that of the ground state. 
Crossed squares are from super-coarse unquenched ensembles, 
closed squares from coarse unquenched ensembles and open squares from the 
fine unquenched ensembles. The closed and open triangles are from the 
coarse and fine quenched ensembles respectively. 
The closed diamond is from the coarse $n_f=2$ ensemble. 
The burst represents the current experimental result.}
\label{fig:leprat}
\end{figure}

For the leading order current only it is easy to extract the matrix 
element from our 
existing results with no further work. 
It is simply the amplitude in the $n$th $\Upsilon$ state of the local 
(delta) smearing, i.e. $a(loc,n)$ from equation~\ref{eq:fit},
appropriately normalised.  
In potential model language this is known as the wavefunction at the 
origin, $\Psi(0)$, and is obtained from the lattice 
results when the correlators are divided by the product 
of the number of colors (3) and the number of non-relativistic 
spins (2). Then $\Psi(0) = \langle \Upsilon_n | J_v | 0 \rangle / \sqrt{6}$, 
accounting for the factor of 6 in 
the denominator of Equation~\ref{eq:gammaeepsi}. 
Results for this quantity are given in 
lattice units in tables~\ref{tab:wfnsuperc},~\ref{tab:wfncoarse} and ~\ref{tab:wfnfine} 
for the super-coarse, coarse and fine ensembles respectively. 

There will be an $\cal{O}$$(\alpha_s)$ systematic error from 
$Z_{match}$ if we try to calculate $\Gamma_{e+e-}$ directly 
from these numbers. Instead we form the ratio: 
\beq
\frac{\Gamma_{ee}(2S)M^2_{\Upsilon(2S)}}{\Gamma_{ee}(1S)M^2_{\Upsilon(1S)}}=\frac{|\Psi_2(0)|^2}{|\Psi_1(0)|^2}
\label{eq:gammratio}
\eeq
in which the $Z_{match}$ factor for the current cancels. We then expect to be 
able to determine this ratio from the leading current with errors 
coming from the matrix elements of relativistic 
corrections to the current.  

Figure \ref{fig:leprat} shows this ratio 
for our lattice results on super-coarse, coarse and fine ensembles. 
The current experimental result~\cite{pdg} is given by the burst. 
The lattice results in the quenched approximation are in strong 
disagreement with experiment. Previous quenched results also 
indicated this, but less precisely~\cite{upspaper,scaling}.  
The results move down towards experiment once light quark vacuum polarization
effects are included and the light quark mass falls. 
However, the results on the unquenched ensembles with lightest 
light quark mass are still some 
way from experiment. The plot shows no dependence on light 
quark mass for the most chiral ensembles, as expected. 
We therefore do not attempt a chiral extrapolation 
but now look at lattice spacing dependence 
for those results with lightest light quark mass.  

\begin{figure}[h]
\includegraphics[width=7cm,angle=-90]{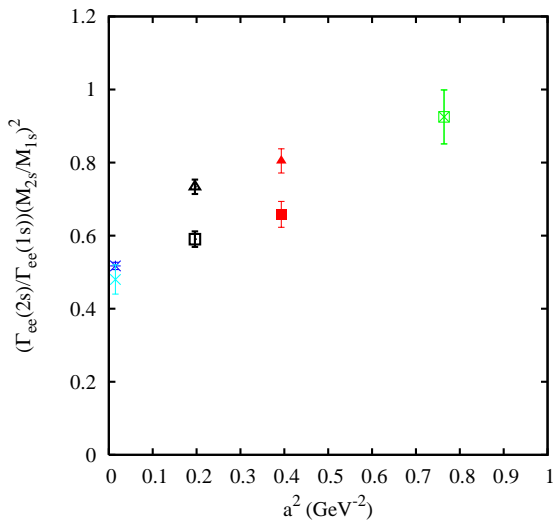}
\caption{The dependence on $a^2$ of the ratio of leptonic width times mass
squared  
for the 1st excited state of the $\Upsilon$ to that of the ground state. 
Crossed, closed and open squares are from the most chiral of 
the super-coarse, coarse and fine unquenched lattices respectively and closed 
and open triangles are
from quenched lattices. 
The burst represents the current experimental result, and the 
cross below this is our continuum extrapolated result, including 
an estimate of the leading relativistic corrections as 
described in the text.}
\label{fig:leprat_vsasq}
\end{figure}

The quenched results for the ratio and those from the most chiral unquenched
ensembles are plotted against the square of the lattice spacing 
in Figure~\ref{fig:leprat_vsasq}. This plot shows a significant 
dependence on $a^2$, indicating discretisation errors in 
the results. These may come from the current operator or from 
the action at the level of $\alpha_sa^2$ and it is not clear 
which is the most important at this stage. The wavefunction at 
the origin is a very short-distance quantity (akin to the 
hyperfine splitting) and it is reasonable 
for it to show significant discretisation effects when other 
quantities, such as radial and orbital splittings in the spectrum, 
do not. 

We can directly calculate the leptonic width for the $\Upsilon$ 
and $\Upsilon^{\prime}$ in the absence of $Z_{match}$ to see 
how each behaves. We use the formula given in Equation~\ref{eq:gammaeepsi}
taking $\alpha_{em}(m_b)$ = 1/132~\cite{erler}. We obtain 
a value $\Gamma_{ee}$=1.43(7) keV for the $\Upsilon$ from 
the super-coarse 2+1 flavor unquenched ensemble,
1.32(1) keV from 
the coarse 2+1 flavor 0.01/0.05 unquenched configurations and 
1.28(1) keV from the fine 2+1 flavor 0.0062/0.031 configurations. The error 
given here comes only from the error in the determination of the ($2S-1S$) 
lattice spacing 
which appears cubed in this result. 
The fine lattice numbers is slightly lower than
the recent preliminary experimental result, 1.34(2) keV~\cite{cleolw}. The 
absence of $Z_{match}$ or higher order corrections gives a 
sizeable additional systematic error to the lattice result, but there 
is at least nothing to indicate that large systematic shifts 
are required to match experiment. The quenched results for 
the $\Upsilon$ are significantly lower at 1.0 keV on the fine 
lattices. For the $\Upsilon^{\prime}$ there is more sign 
of discretisation in the bare leading-order matrix element. 
From this, we obtain $\Gamma_{e+e-}$ = 1.2(1) keV on the 
super-coarse unquenched ensemble, 0.77(5) keV on the 
coarse unquenched lattices and 0.67(3) keV on the fine unquenched
lattices, with errors as above. 
This 
is not inconsistent with the fact that the excited 
state has more structure in its `wavefunction' and therefore 
might be more susceptible to short distance errors than the 
ground state. 
The fine lattice result is slightly higher than the recent preliminary experimental value of 
0.616(13) keV~\cite{cleolw}. 

As for the hyperfine splitting earlier (Equation~\ref{eq:hypvsa})
we consider $a$-dependence in the ratio of Equation~\ref{eq:gammratio}
of the form
\begin{equation}
{\rm ratio}(a) = {\rm ratio}(0)(1+(\Lambda a)^2 + \ldots).
\label{eq:ratvsa}
\end{equation}
However, here the $a$-dependence is stronger than for the 
hyperfine splitting ($\Lambda \approx$ 900 MeV) and the super-coarse results are therefore 
correspondingly less reliable. Using again a constrained 
fit we 
find a continuum value for the ratio of 
0.50(4) for the leading order current. The 8\% error 
is a combination of the statistical/fitting error and 
the effect of discretisation errors. 

The higher order relativistic current corrections are given 
by operators of the form~\cite{joneswol}:
\begin{equation}
\chi^{\dag}(\frac{1}{8(M_b^0)^2}(\Delta^{(2)\dag}\sigma + \sigma\delsq - 
2(\sigmav\cdot\delv^{\dag})\sigma(\sigmav\cdot\delv))\psi. 
\end{equation}
For an $\Upsilon$ at zero momentum the $\delsq$ operator acting 
on either quark or anti-quark will give the same matrix element. 
This is similarly true for the operator made from one derivative 
on each. The $i$th component of the correction operator then becomes:
\begin{equation}
\chi^{\dag}\frac{1}{4(M_b^0)^2}(\frac{2}{3}\Delta^{(2)\dag}\sigma_i 
+ 2D_{ij}\sigma_j)\psi. 
\end{equation}
where $D_{ij}=\del_i\del_j - \delsq/3$. The second term above 
is the piece which couples to a $D$-wave state~\cite{upspaper} and 
gives a leptonic width to the $^3D_1$ state. The first term 
is the one that we are concerned with and that gives a relativistic 
correction to the $S$-wave leptonic width of the form 
$\chi^{\dag}\delsq\sigma/6(M_b^0)^2\psi$. We calculate the 
matrix element of this term by inserting a $\delsq$ operator 
at the correlator sink. We then fit the correlator with the insertion 
simultaneously with the other correlators used to get the 
matrix element of the leading order current above. 
We have done this for the 0.01/0.05
2+1 flavor coarse ensemble only. There the 
current correction matrix element (in lattice units) 
is -0.0107(2) for the $\Upsilon$ and -0.0131(4) for the 
$\Upsilon^{\prime}$, i.e. a 3.5\% and 5.4\% correction to the 
leading matrix element respectively. This induces a further 4\% 
reduction in the ratio of leptonic width times mass squared 
for the 2$S$ to 1$S$. The size of the relativistic 
corrections and the overall shift from their difference 
is perfectly in keeping with expectations based on 
power-counting in $v^2$. Higher order relativistic corrections 
are then likely to be negligible. 

We apply this shift of 4\% to our continuum leading 
order current result. We also add an additional 4\% 
error ($\alpha_sv^2$) to allow for missing radiative 
corrections to the relativistic corrections in the 
ratio (radiative corrections to the leading order term 
cancel, as discussed earlier). This gives our final answer for 
the ratio 
of Equation~\ref{eq:gammratio} 
of 0.48(5), to be compared to the recent preliminary experimental 
result~\cite{cleolw} of 0.517(10). 
Both of these are also marked on Figure~\ref{fig:leprat_vsasq}. 
Our result is compatible with experiment but not as precise. 
Our 10\% error is dominated 
by the fact that the lattice results depend strongly on the 
lattice spacing.  

The lattice calculation of $\Gamma_{ee}$ can be significantly 
improved. 
Once the renormalisation of the leading and sub-leading lattice NRQCD current 
operators is calculated, we will be able to compute the $\Gamma_{ee}$
results for each radial excitation of the $\Upsilon$ separately 
with systematic errors at the 10\% level 
coming from unknown $\alpha_s^2$ terms in $Z_{match}$~\cite{horgan}. 
We will also discover whether the current corrections contain a 
sizeable contribution in the form of a discretisation 
correction, so that the scaling with lattice spacing improves. 
As discussed earlier, the ratio of leptonic width times mass squared 
will be more precise than this because 
the errors will be set by $v^4$ corrections to the current and 
$\alpha_s^2$ errors in the $v^2$ current correction piece. 
These errors should be at the level of a few percent. 

\section{Conclusion}\label{sec:conc}

\begin{figure}[t]
\includegraphics[width=7cm,angle=-90]{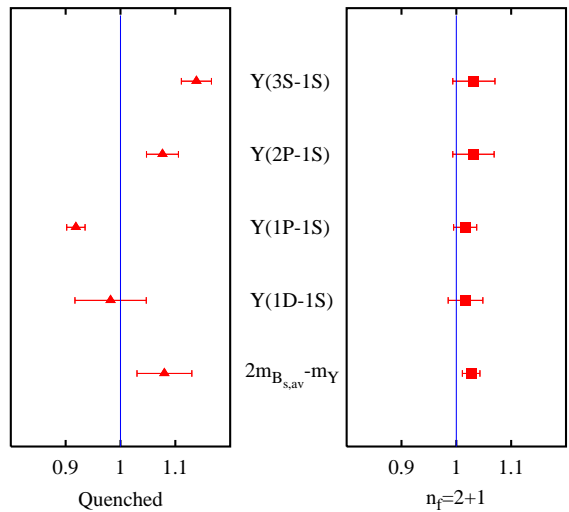}
\caption{The ratio of lattice results to experiment for various 
mass splittings involving $b$ quarks. 
The results are for quenched simulations (left) and 
simulations incorporating $2+1$ flavors of sea
quarks (right). All the unquenched results come 
from this work. We have taken results from the most 
chiral coarse lattice ensemble since those have the 
best errors in general. Errors include statistical/fitting 
errors for the quantity in question and those from 
determining the lattice spacing. 
The top three quenched results come from this work, but 
the lower two are estimates from earlier work.  
The quenched $1D-1S$ splitting is taken from~\cite{campbelm} using 
the $^1D_2$ state. The quenched $2m_{B_s}-m_{\Upsilon}$ splitting is 
estimated from~\cite{joachim}, adjusted to use 
the lattice spacing from the $\Upsilon$ $2S-1S$ 
spacing. 
}
\label{newratio}
\end{figure}

We have presented a state-of-the-art calculation for
$\Upsilon$ spectroscopy in lattice QCD. The major 
improvement over previous calculations is that we 
now have results on gluon field configurations that 
include 2+1 flavors of sea quarks with the $u/d$ 
quark masses within reach of their physical values. 
This means, as we demonstrate here, that it is now 
possible to fix the parameters of the QCD Lagrangian 
unambiguously. The $\Upsilon$ system is a good 
place to test this because of the large number of 
well-determined excited states. Figure~\ref{newratio} 
shows a `ratio' plot of lattice results from this 
paper compared
to experiment. The unquenched results on the right are 
in striking agreement with experiment compared to 
the quenched results on the left. This ratio plot 
differs slightly from that in~\cite{us} for 
the quenched results because of improved fits for 
the quenched ensembles here. 

Our results enable us to determine 
physical values for the parameters $r_0$ (0.469(7) fm) 
and $r_1$ (0.321(5) fm),
often used to compare lattice spacing values between 
different ensembles in lattice calculations. 

Further improvements to the calculation of 
radial and orbital excitation energies will 
require further improvements to the NRQCD action, in 
particular the calculation of radiative corrections 
to $v^4$ and $a^2$ terms. This is in progress. Statistical 
precision can also be improved since double 
the number of configurations used here are available
as well as further configurations with lighter sea
quark masses~\cite{milc1, milc2}. Precision on 
$P$ and $D$-wave states would be improved by better 
smearing.  

Our result for the $b$ quark mass in the $\overline{MS}$ scheme 
at its own scale is 4.4(3) GeV, using a 1-loop matching of 
the lattice to the continuum.
A 2-loop calculation of the heavy quark self-energy would reduce the 
error significantly on the $b$ quark mass as well as make the 
remaining error estimate more reliable by fixing the scale for 
$\alpha_s$ in the matching coefficient. This is currently underway.  

The fine structure in the spectrum is only calculated to 
leading order here, so sizeable systematic errors remain. 
We are able nevertheless to predict the ground-state 
hyperfine splitting more accurately than in the past 
and the radially excited hyperfine splitting for the 
first time. Our results are 61(14) MeV and 30(19) MeV 
respectively. 
To reduce the errors requires, as above, the determination of radiative 
corrections to the leading spin-dependent terms in the NRQCD  
action and the inclusion of sub-leading terms. Further 
improvement to reduce discretisation errors is also an 
issue. 

The calculation of the leptonic width for $\Upsilon$ and 
$\Upsilon^{\prime}$ is an important example of a 
simple decay rate that can be calculated in lattice QCD, 
without the ambiguity that arises from potential 
models~\cite{eq}. Our results on unquenched configurations for the ratio of 
leptonic widths (multiplied by squared masses) for $\Upsilon^{\prime}$ and $\Upsilon$ 
show a much closer agreement with experiment than 
those on quenched configurations. Sizeable discretisation 
errors prevent us giving a very accurate result here, 
but we obtain 0.48(5) for the ratio. 
Further precision 
will come from the calculation of the renormalisation 
constant required to match the lattice and 
continuum currents responsible for the decay. 
The calculation will yield an improved current operator 
with reduced discretisation errors. 
This is also underway. 

\vspace{5mm}
{\bf{Acknowledgements}}
We thank members of the MILC collaboration for making their configurations 
available, and R. Sugar, D. Toussaint and S. Gottlieb 
for help with reading them. We thank R. Horgan, Q. Mason, M. Nobes, J. Pivarski and 
H. Trottier 
for useful discussions. The computing was done on the 
UKQCD alpha cluster at the University of Glasgow funded by 
PPARC and on the Fermilab cluster and at NERSC
funded by the U.S. Department of Energy. We are grateful to 
PPARC, the DoE and the NSF for support.


\begin{thebibliography}{99}
\bibitem{comment} The value of $u_0$ does affect the value of the 
ground state energy as is clear from comparing Tables~\ref{tab:SI_E_coarse} 
and~\ref{tab:SI_sys}. The dependence on $u_0$ can be seen in perturbation theory 
or in a mean field approach~\cite{nakhleh} which gives a 
reasonably good quantitative description of the results.
\end{thebibliography}
\end{document}